\documentclass[referee,usenatbib,useAMS]{mn2e}
\usepackage{times}
\usepackage{multirow}
\usepackage{color}
\usepackage{multicol}
\usepackage{graphicx}
\usepackage{float}
\usepackage{pdflscape}
\usepackage{threeparttable}
\usepackage{epstopdf}
\usepackage{amsmath}
\usepackage{amssymb}
\title[Short term X-ray variability of \pds]{Short term X-ray spectral variability of the quasar \pds\ observed in a low flux state}
\author[Matzeu et al.]
{G. A. Matzeu,$^{1}$\thanks{Correspondence to: g.matzeu@keele.ac.uk} 
J. N. Reeves, $^{1,2}$ E. Nardini, $^1$ V. Braito, $^{2,3}$ M. T. Costa, $^1$
\newauthor
F. Tombesi, $^{4,5}$ 
J. Gofford, $^1$ \\
$^1$Astrophysics Group, School of Physical and Geographical Sciences, Keele University, Keele, Staffordshire ST5 5BG, UK\\
$^2$Center for Space Science and Technology, University of Maryland Baltimore County, 1000 Hilltop Circle, Baltimore, MD 21250, USA\\
$^3$INAF – Osservatorio Astronomico di Brera, Via Bianchi 46, I-23807 Merate (LC), Italy\\
$^4$X-ray Astrophysics Laboratory, NASA/Goddard Space Flight Center, Greenbelt, MD, 20771, USA\\
$^5$Department of Astronomy and CRESST, University of Maryland, College Park, MD, 20742, USA}

\newcommand{\pds}{PDS\,456 }
\newcommand{\xmm}{{\it XMM-Newton }}

\newcommand{\suzaku}{{\it Suzaku }}

\newcommand{\xstar}{\textsc{xstar }}
\newcommand{\Msun}{\hbox{$\thinspace M_{\odot}$}}

\newcommand{\nustar}{\textit{NuSTAR }}

\begin{document}

\date{\today}

\pagerange{\pageref{firstpage}--\pageref{}} \pubyear{?}

\maketitle
\label{firstpage}
\begin{abstract}

\noindent We present a detailed analysis of a recent, 2013 \suzaku campaign on the nearby ($z=0.184$) luminous (L$_{\rm bol}\sim10^{47}$ erg s$^{-1}$) quasar PDS 456. This consisted of three observations, covering a total duration of $\sim1$ Ms and a net exposure of $455$ ks. During these observations, the X-ray flux was unusually low, suppressed by a factor of $>10$ in the soft X-ray band when compared to previous observations. We investigated the broadband continuum by constructing a Spectral Energy Distribution (SED), making use of the optical/UV photometry and hard X-ray spectra from the later simultaneous \xmm and \nustar campaign in 2014. The high energy part of this low flux SED cannot be accounted for by physically self consistent accretion disc and corona models without attenuation by absorbing gas, which partially covers a substantial fraction of the line of sight towards the X-ray continuum. At least two layers of absorbing gas are required, of column density $\log (N_{\rm{H,low}}/{\rm cm^{-2}})=22.3\pm0.1$ and $\log (N_{\rm{H,high}}/{\rm cm^{-2}})=23.2\pm0.1$, with average line of sight covering factors of $\sim80\%$ (with typical $\sim5\%$ variations) and $60\%$ ($\pm10-15\%$), respectively. During these observations \pds displays significant short term X-ray spectral variability, on timescales of $\sim 100$\,ks, which can be accounted for by variable covering of the absorbing gas along the line of sight. The partial covering absorber prefers an outflow velocity of $v_{\rm pc} = 0.25^{+0.01}_{-0.05}c$ at the $>99.9\%$ confidence level over the case where $v_{\rm pc}=0$. This is consistent with the velocity of the highly ionised outflow responsible for the blueshifted iron K absorption profile. We therefore suggest that the partial covering clouds could be the denser, or clumpy part of an inhomogeneous accretion disc wind. Finally estimates are placed upon the size-scale of the X-ray emission region from the source variability. The radial extent of the X-ray emitter is found to be of the order $\sim 15-20~R_{\rm g}$, although the hard X-ray ($>2$ keV) emission may originate from a more compact or patchy corona of hot electrons, which is typically $\sim6-8~R_{\rm g}$ in size.


\end{abstract}


\section{Introduction}

\noindent Outflows are now considered an essential component in the overall understanding of Active Galactic Nuclei (AGN). These winds are thought to occur as a result of the accretion process \citep{King03,Ohsuga09}, and they can provide a link between the black hole mass and the velocity dispersion of the stars in the bulge of a galaxy, such as seen with the $M-\sigma$ relation \citep{Ferrarese00,Gebhardt00}. Several works have tried to understand this relation in terms of the negative mechanical feedback provided by winds produced by an accreting super-massive black hole \citep[e.g.,][]{King03,King10,DiMatteo05,McQuillin13}. A number of massive and high velocity outflows have been detected in luminous AGN \citep{Chartas03,Reeves03,Pounds03} through the presence of resonance iron K-shell absorption lines blue-shifted to energies of $E>7$ keV (in the rest-frame).
\noindent The importance of these winds is supported by their frequent detection, as they are observed in the X-ray spectra of approximately $40\%$ of AGN \citep{Tombesi10,Gofford13}, suggesting that their geometry is characterised by a wide solid angle. This was recently confirmed in the quasar \pds by \citet[][hereafter N15]{Nardini15}. These fast outflows are characterised by a considerably high column density ($N_{\rm H}\sim 10^{23} $cm$^{-2}$) and a mean velocity $\left \langle v_{\rm w} \right \rangle\sim0.1c$ \citep{Tombesi10}. However, the velocity of the Fe K absorbers can cover a wide range, from as low as a few $\times100-1000$ km s$^{-1}$ (more typical of what is seen in the soft X-ray warm absorbers; \citealt{Kaastra00,Blustin05,McKernan07}) up to mildly relativistic values of $\sim0.2-0.4c$ in the more extreme cases \citep[e.g.,][hereafter R09]{Chartas02,Reeves09}. These high velocities can result in a large amount of mechanical power, possibly exceeding the $0.5-5\%$ of the bolometric luminosity $L_{bol}$ required for a significant AGN feedback contribution to the evolution of the host galaxy \citep{King03,KingP03,DiMatteo05,HopkinsElvis10,Tombesi15}.
\\
\noindent The primary X-ray emission of AGN is thought to originate from \textquotedblleft seed" UV disc photons that are Compton (up)scattered in a corona of relativistic electrons \citep{HaardtMaraschi91,HaardtMaraschi93}, producing the hard X-ray tail usually phenomenologically described by means of a simple power law. The AGN X-ray variability could be characterised by either intrinsic fluctuating spectral and temporal behaviour \citep[e.g.][]{Marshall81,Green93} or the presence of absorbing gas in the line of sight \citep{Risaliti09b}. The latter case may be seen as changes in the covering fraction of a partial covering absorber \citep[e.g.][]{Turner11}. This may favour in particular the explanation of the X-ray variability of several AGN, whose low flux and hard broadband spectra are possibly due to partial occultation by absorbing clouds \citep[e.g.,][]{Risaliti05,Turner08,Behar10}. Since the last decade, partial covering scenarios have been successful in explaining the complex X-ray spectral properties of AGN in different energy bands such as: pronounced continuum curvature below $10$ keV \citep[e.g.,][]{Miller08}, rapid spectral variability \citep[e.g.,][]{NardiniRisaliti11}, X-ray occultation \citep[e.g.,][]{Risaliti05b}, and pronounced hard excesses above $10$ keV  \citep[e.g.,][]{Turner09,Tatum13}. In the latter case, however, reflection models have also been invoked to explain some sources with strong hard excesses \citep[e.g.,][]{Nardini11,Risaliti13}.
\\  
\noindent The luminous radio-quiet quasar \pds is located at a redshift of $z=0.184$ \citep{Torres97}, and it has a de-reddened absolute magnitude of M$_B\sim-27$ and a bolometric luminosity of $L_{bol}\sim10^{47}$\,erg\,s$^{-1}$ \citep{Simpson99,Reeves00}. It is comparable in luminosity to the radio-loud quasar 3C 273, making it the most luminous quasar in the local Universe ($z<0.3$). Such a high luminosity is more typical of quasars at redshift $z=2-3$, considered the peak of the quasar epoch, where black hole feedback was thought to play a key role in the evolution of galaxies \citep{DiMatteo05}. The extreme X-ray nature of \pds was first noticed by \citet{Reeves00}, where very rapid X-ray variability, on time-scales of $\sim15$ ks, was observed from RXTE monitoring observations in the $3 - 10$ keV band. This indicates, by the light-crossing time argument, a very compact X-ray source of a few gravitational radii ($R_{\rm g}$) in extent (where $R_{\rm g}=GM_{\rm BH}/c^2$). A short ($40$ ks) observation carried out with \xmm in 2001 detected a strong absorption trough in the iron K band, above $7$ keV, possibly attributed to the highly ionised iron K-shell feature with an associated outflow velocity of $v_{\rm w}\gtrsim 0.1c$ \citep{Reeves03}. A longer ($190$ ks) 2007 \suzaku observation \textquotedblleft cemented" the evidence for this fast outflow, revealing two highly significant absorption lines centred at $9.08$ and $9.66$ keV in the quasar rest frame, where no strong atomic transitions are otherwise expected. The association of these lines to the nearest expected strong line, the Fe\,\textsc{xxvi} Ly$\alpha$ transition at $6.97$ keV, implied an outflow velocity of $\sim0.25-0.30c$ (R09). Similarly, in a more recent (2011) $120$ ks \suzaku follow-up observation, a broad absorption trough at $\sim9$ keV (in the source rest frame) was again found, confirming that in both the 2007 and 2011 observations the changes in the absorption features could be due to the same flow of gas in photo-ionisation equilibrium with the emergent X-ray emission \citep[][hereafter R14]{Reeves14}. Furthermore, in a recent series of five simultaneous observations with \xmm and \nustar in $2013-2014$, N15 resolved a fast ($\sim0.25c$) P-Cygni like profile at Fe K, showing that the absorption originates from a wide angle accretion disc wind. Indeed, \pds has a proven track record of strong X-ray spectral variability over the last decade, likely due to absorption and intrinsic continuum variations \citep{Behar10}. 
\\
\noindent Here we present a long ($\sim1$ Ms duration) 2013 \suzaku campaign carried out in order to determine the timescales through which both the X-ray absorption and continuum variations occur, by directly measuring the absorber's behaviour on timescales of tens of ks (corresponding to a light-crossing time of a few $R_{\rm g}$ for $M_{\rm BH}\sim10^{9}M_{\odot}$). An initial analysis of this dataset has been presented in \citet[][ hereafter G14]{Gofford14}, focusing on the variability of iron K absorption feature. In this paper we investigate in more detail the broadband continuum and absorption variability.

\begin{table*}

\centering
\begin{tabular}{cccccc}

\hline



&2007&2011&2013a&2013b&2013c\\

Obs.~ID&$701056010$&$705041010$&$707035010$&$707035020$&$707035030$\\

Start Date, Time (UT)&2007-02-24, 17:58&2011-03-16, 15:00&2013-02-21, 21:22&2013-03-03, 19:43&2013-03-08, 12:00\\

End Date, Time (UT)& 2007-03-01, 00:51& 2011-03-19, 08:33& 2013-02-26, 23:51& 2013-03-08, 12:00&2013-03-11, 09:00\\

Duration(ks)& 370 & 240 & 440.9 & 404.2 & 248.4 \\

Exposure(ks)$^{a}$& 190.6& 125.6 & 182.3 & 164.8 & 108.3 \\

\hline

Flux$_{(0.5-2)\rm keV}^{b}$& 3.46 & 1.36 & 0.59 & 0.30 & 0.43 \\

Flux$_{(2-10 )\rm keV}^{b}$&3.55 & 2.84 & 2.09 & 1.59 & 1.72 \\


Flux$_{(15-50)\rm keV}^{b}$&$5.7_{-2.2}^{+2.2}$& $< 2.5$ & $<2.0$ & $<2.0$ & $<2.0$ \\

\hline
\

\end{tabular}
\caption{Summary of the three 2013 observations of \pds with Suzaku plus the 2007 and 2011 for comparison purposes.}
\vspace{-5mm}
\begin{threeparttable}
\begin{tablenotes}
	\item[a] Net Exposure time, after background screening and dead-time correction.
	\item[b] Observed fluxes in the $0.5 - 2$\,keV, $2 - 10$ \,keV and $15 - 50$\,keV bands in units $\times10^{-12}$\,erg cm$^{-2}$ s$^{-1}$ ($1\sigma$ upper limits).	
	
\end{tablenotes}
\end{threeparttable}
\label{tab:suzaku}
\end{table*}

\begin{table*}

\centering
\begin{tabular}{ccccc}

\hline

&\textit{XMM}/ObsA&\textit{XMM}/ObsE&\textit{NuSTAR}/ObsA&\textit{NuSTAR}/ObsE\\
\hline    

Obs.~ID&$0721010201$&$0721010601$&$60002032002$&$60002032010$\\

Start Date, Time (UT)&2013-08-27, 04:41&2014-02-26, 08:03&2013-08-27, 03:41&2014-02-26, 08:16\\

End Date, Time (UT)&2013-08-28, 11:13&2014-02-27, 22:51&2013-08-28, 11:41&2014-02-28, 22:56\\

Duration(ks)& 110.0 & 139.7 & 113.9 & 224.3  \\

Exposure(ks)& 95.7& 100.3 & 43.8 & 109.7  \\

\hline

Flux$_{(0.5-2)\rm keV}$      & 3.8     & 1.6      & --    & --  \\

Flux$_{(2-10)\rm keV}$      &6.6      & 2.6      & 6.8   & 2.7  \\

Flux$_{(15-50)\rm keV}$     & --      & --       &3.9    & 1.2  \\

\hline
\

\end{tabular}
\caption{Summary of the simultaneous \xmm and \nustar observations of \pds focused on observations A and E. Quantities and units are the same defined in Table 1.}
\vspace{-5mm}

\label{tab:xmm/nustar}
\end{table*}

\section{Data Reduction}

In this work the events and spectra adopted are essentially the same as in G14, with the only difference being a further improvement on the constraint on the iron K absorption profile. This was achieved by re-binning the spectra to the Half-Width at Half-Maximum (HWHM) energy resolution of the detector (i.e., $\sim60$ eV at $6$ keV).
\\ 
\suzaku \citep{Mitsuda07} observed \pds between February and March 2013 through the X-ray Imaging Spectrometer \citep[XIS;][]{Koyama07} and the Hard X-ray Detector \citep[HXD;][]{Takahashi07}, although in these observations \pds was not detected in either the PIN or GSO detectors. Due to scheduling reasons the observation is constituted by three sequences (see Table~\ref{tab:suzaku}): the first (OBSID:707035010, hereafter 2013a), obtained between the 21-26 of February 2013, has a duration of $\sim441$ ks; the second (OBSID:707035020, hereafter 2013b) and the third (OBSID:707035030, hereafter 2013c) were obtained consecutively between the 3-11 of March 2013, and have durations of $\sim404$ ks and $\sim248$ ks, respectively. The total duration of the campaign was $\sim1$ Ms (after allowing for a scheduling gap between the first and the second sequence), with a total net exposure of $455$ ks (Table~\ref{tab:suzaku}). For completeness, the observation details for the earlier 2007 and 2011 \suzaku observations, when the X-ray flux was substantially higher, are also listed in Table~\ref{tab:suzaku}. All the spectral analysis and model fitting, performed with XSPEC v12.8.2 \citep{Arnaud96}, were focused on the spectra obtained by the XIS front illuminated (FI) CCDs --- XIS 0 and XIS 3 --- as they are characterised by a larger effective area and yet lower background in the iron K band compared to the back illuminated XIS 1 CCD. The XIS 1 spectra, although consistent with those obtained with XIS 0 and 3, are noisier at higher energies in the Fe K band. 
\\
Furthermore, as the XIS 0 and XIS 3 spectra were consistent with each other, we combined them into a single XIS-FI spectrum for all the observations. For the XIS-FI spectra we adopted a spectral binning corresponding to the approximate HWHM resolution of the detector (which is $\sim 60$\,eV at 6\,keV), using the \texttt{rbnpha} and \texttt{rbnrmf} \textsc{ftools} to hardwire this binning into the spectral and response files. An additional grouping corresponding to $>25$ counts per spectral bin was subsequently applied to the rebinned spectra, in order to use the $\chi^2$ minimization technique. The XIS-FI spectra are fitted over the $0.6-10$ keV band, ignoring the $1.7-2.1$ keV interval due to uncertainties with the XIS detector Si edge. Although not detected at hard X-rays in these observations, we obtained an upper limit to the HXD/PIN flux in 2013a between $15-50$ keV, i.e. $<2.0\times10^{-12}$ erg cm$^{-2}$ s$^{-1}$. The flux was obtained by converting the HXD/PIN count rate through WebPIMMS\footnote{https://heasarc.gsfc.nasa.gov/cgi-bin/Tools/w3pimms/w3pimms.pl}, assuming a simple power law with $\Gamma=2.4$ (see below). In all the subsequent spectral fits, the predicted limits to the hard X-ray flux are consistent with the above value.
\\
Values of $H_0=70$ km s$^{-1}$ Mpc$^{-1}$ and $\Omega_{\Lambda_{0}}=0.73$ are assumed throughout and errors are quoted at the $90\%$ confidence level ($\Delta \chi^{2}=2.71$) for one parameter of interest.


\section{broadband Spectral Analysis}

\begin{figure}
\begin{center}

\includegraphics[width=10cm]{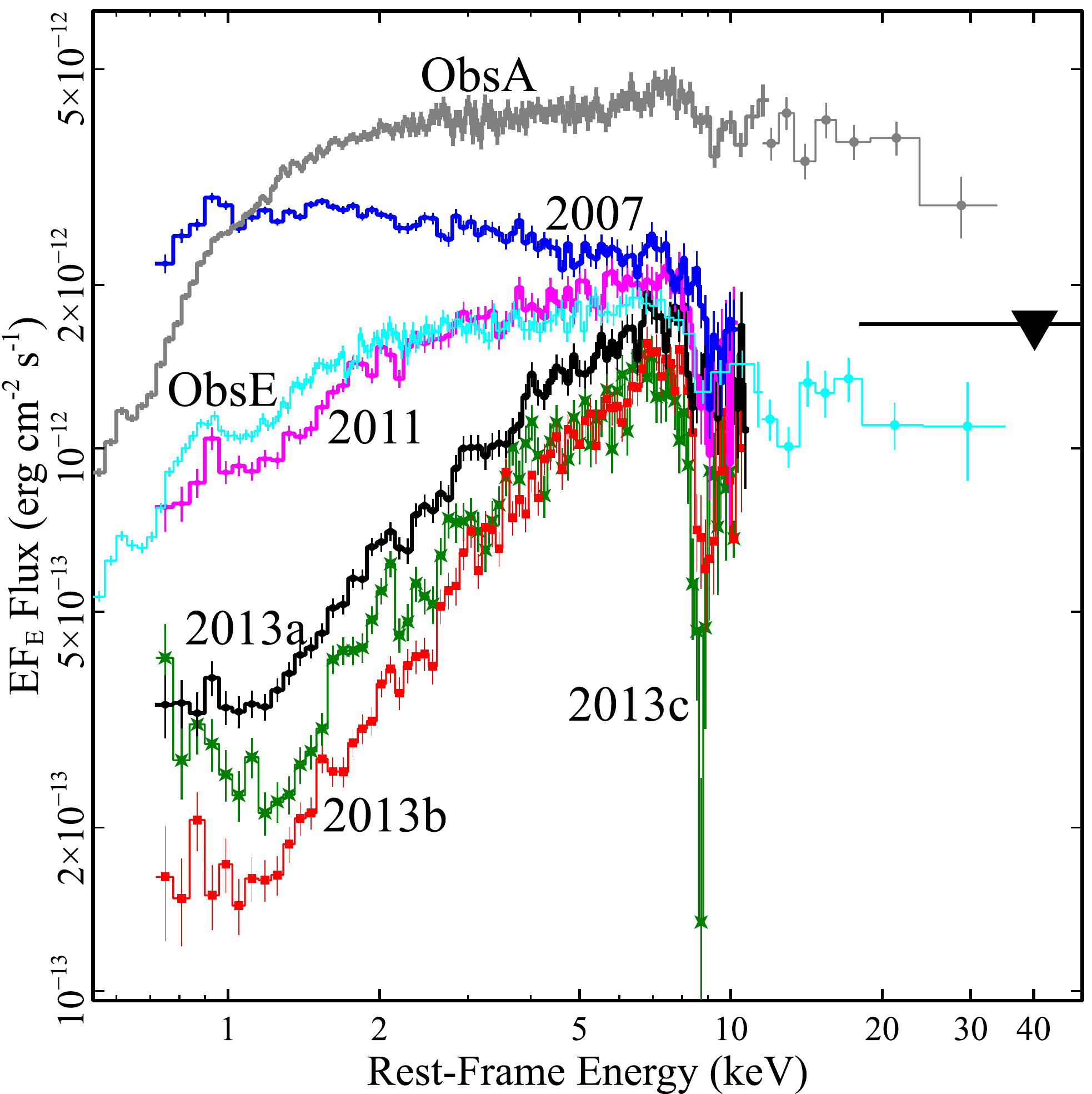}	
\caption{Spectra of the 2013 \suzaku observation: Sequence 1 - 2013a (black, including the $15-50$ keV HXD/PIN upper limit), Sequence 2 - 2013b (red) and Sequence 3 - 2013c (green). For comparison also the 2013 (Obs A, grey) and 2014 (Obs E, cyan) spectra of the {\xmm}/{\nustar} campaign and the 2007 (blue) and 2011 (magenta) \suzaku spectra are shown. The long term spectral variations in the soft band and in the iron-K region, which characterise this type 1 quasar, are evident. The \suzaku observation in 2013 caught \pds in a remarkably low flux state compared to both earlier and subsequent observations.}
\label{fig:pds_xmm_nu_suz_good}
\end{center}
\end{figure}

The 2013 \suzaku observations caught \pds in an unusually low flux, compared to either the later simultaneous \xmm and \nustar campaign, carried out in August 2013/February 2014, and to the 2007 and 2011 \suzaku observations (Tables~\ref{tab:suzaku} and \ref{tab:xmm/nustar}). This is shown in Fig.~\ref{fig:pds_xmm_nu_suz_good}, which compares the fluxed spectra (unfolded through the instrumental response against a simple $\Gamma=2$ power law, and not corrected for Galactic absorption) from the 2013 \suzaku sequences to the lowest (Obs E) and the highest (Obs A) of the five \xmm/\nustar sequences (N15). It is clear that the 2013 \suzaku observations caught PDS 456 in an extended period of low flux, especially compared to the 2007 \suzaku or the 2013 \xmm (Obs A) observations, when the soft X-ray band flux is typically a factor of $10$ brighter. This therefore provided a unique opportunity to understand the properties of the intrinsic continuum and the reprocessing material in the low flux state of PDS 456. 

\begin{table*}

\begin{tabular}{cc|ccc|}
\hline\hline
Component                                &Parameter                                &2013a                 &2013b                         &2013c    \\
\hline

\multirow{1}{*}{\texttt{Tbabs}}    &$N_{\textsc{h},\rm{Gal}}$/cm$^{-2}$                &     $2\times10^{21}$                  &$2\times10^{21}$     &   $2\times10^{21}$       \\
\\

\multirow{6}{*}{\texttt{optxagnf}}       &$\log(L/L_{\rm Edd})$                    &$-0.08\pm0.10$         &$-0.08^{t}$                 &$-0.08^{t}$\\

                                         &r$_{\rm cor}$ (R$_{\rm g}$)              &$41_{-19}^{+58}$      &$41^{t}$                    &$41^{t}$    \\

                                         &kT$_{\rm e}$ (eV)                        &$468_{-156}^{+156}$   &$468^{t}$                   &$468^{t}$   \\

										&$\tau$                                  &$9.9_{-2.4}^{+2.3}$    &$8.9_{-2.7}^{+2.5}$         &$9.6\pm2.7$ \\

										&$\Gamma$                                &$2.4\pm0.1$             &$2.4^{t}$                   &$2.4^{t}$  \\

										&F$_{\rm pl}$                         &$0.09_{-0.04}^{+0.07}$      &$0.08_{-0.03}^{+0.06}$       &$0.11_{-0.04}^{+0.08}$ \\

\\
\multirow{2}{*}{\texttt{pc}$_{\rm low}$} &$\log$($N_{\rm{H,low}}$/cm$^{-2}$)   &$22.4\pm0.1$   &$22.4^{t}$                           &$22.4^{t}$\\

                                         &f$_{cov,\rm low}$ ($\%$)                 &$88_{-3}^{+2}$         &$88^{t}$                     &$88^{t}$\\
\\

\multirow{2}{*}{\texttt{pc}$_{\rm high}$}&$\log$($N_{\rm{H,high}}$/cm$^{-2}$)  &$23.1\pm0.1$           &$23.1^{t}$                   &$23.1^{t}$\\

                                         &f$_{cov,\rm high}$ ($\%$)                &$71_{-9}^{+10}$         &$71^{t}$                    &$71^{t}$\\

\\

\multirow{3}{*}{\texttt{zgauss$_{em}$}}     &Energy (keV)                             &$6.9\pm0.1$            &$6.9^{t}$                    &$6.9^{t}$\\

                                         &EW (eV)                                  &$94_{-26}^{+32}$       &$118_{-36}^{+33}$            &$125_{-31}^{+43}$\\

                                         &$\sigma$(eV)                             &$212^{t}$              &$212^{t}$                    &$212^{t}$\\

										&norm$^a$                            &$4.6\pm1.2$              &$4.6^{t}$                    &$4.6^{t}$\\

\\

\multirow{4}{*}{\texttt{zgauss$_{abs}$}}

                                         &Energy (keV)                          &$8.55_{-0.26}^{+0.34}$    &$8.88\pm0.10$              &$8.70\pm0.07$\\

                                         &EW (eV)                               &$126\pm63$                &$353\pm82$                 &$530\pm72$\\

                                         &$\sigma$(eV)                          &$264^{\rm t}$             &$264^{\rm t}$              &$264_{-56}^{+74}$\\

										&norm$^a$                            &$-3.0_{-0.76}^{+1.5}$              &$-6.4\pm1.4$                    &$-9.6\pm1.3$\\

                                         &($\Delta \chi^2/\Delta \nu$)$^b$             &$15/2$                   &$61/2$                    &$141/2$\\

\hline

&&$\chi^{2}/{\nu}=522/498$&&\\

\hline
\end{tabular}

\caption{\texttt{optxagnf} model baseline continuum parameters. $^t$ denotes that the parameter is tied during fitting. A cross-normalisation factor of $0.98\pm0.08$ has been found between the non simultaneous \suzaku 2013a and \nustar Obs E spectra.}
\vspace{-5mm}
\begin{threeparttable}
\begin{tablenotes} 
	\item[\small{$L/L_{\rm Edd}$: Eddington ratio}]  
	\item[\small{r$_{\rm cor}$ radius of the X-ray corona in R$_{\rm g}$,}]
	\item[\small{F$_{\rm pl}$: fraction of the dissipated accretion energy emitted in the hard power law,}]
	\item[\small{\texttt{pc$_{\rm low}$}, \texttt{pc$_{\rm high}$}: low and high column partial covering components with respective column density and covering fraction,}]	
 	\item[a] Gaussian emission and absorption profile normalisation, in unit of $10^{-6}$ photons cm$^{-2}$ s$^{-1}$.
 	\item[b] Change in $\Delta\chi^{2}/{\Delta\nu}$ when the Gaussian component modelling the iron K absorption profile is removed. 
\\	
 
\end{tablenotes}
\end{threeparttable}
\label{tab:optxagnf}
\end{table*}

\subsection{Comparison between \textit{Suzaku} and \textit{XMM-Newton/NuSTAR}}

\begin{figure}
\begin{center}
\includegraphics[scale=0.7]{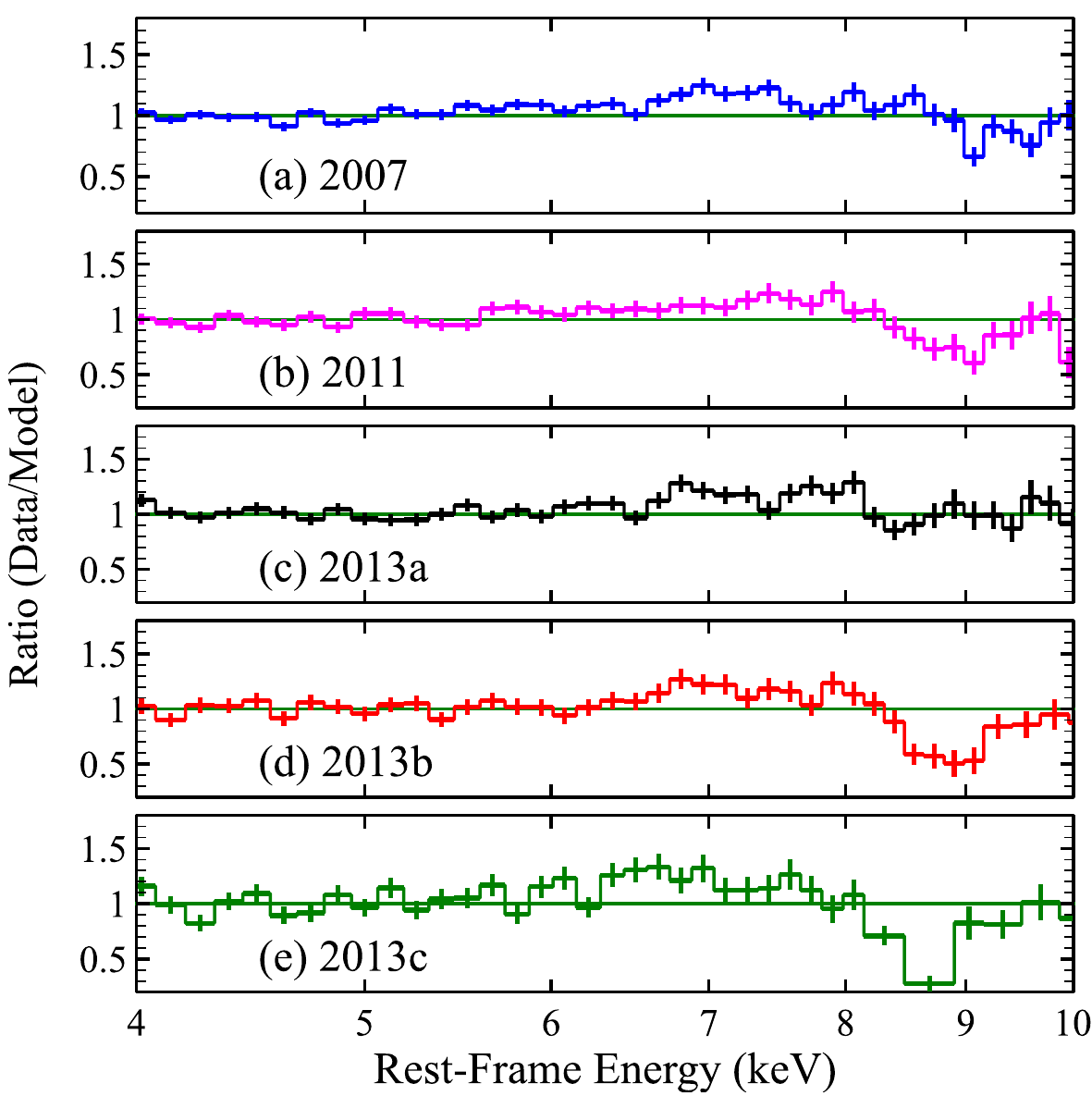}	
\caption{Fe K band of the 2007 (blue), 2011 (magenta), 2013a (black), 2013b (red) and 2013c (green) \suzaku observations plotted as a ratio to a phenomenological continuum fitted by excluding the observed $5-8.5$ keV range. It is noticeable how in the 2013 segments, the strength of the absorption feature increases at energies $E\sim8.5-9$ keV as the observation progresses. The broad iron K emission is rather faint and appears to be centred at $\sim7$ keV, while the narrow Fe K$\alpha$ line at $6.4$ keV is virtually absent from all the five observations.}
\label{pds456_5seq_time_fek}
\end{center}
\end{figure}

As a preliminary test, we parameterised the spectral differences between all the five \suzaku spectra in the $2-8$ keV band (in order to avoid the contribution of the Fe-K absorption lines to the continuum) with a simple power-law model, where the photon indices and normalisations are allowed to vary between all the spectra. The 2011, 2013a, 2013b and 2013c are characterised by a harder spectral shape, $\Gamma=1.94\pm0.04$, $\Gamma=1.47\pm0.04$, $\Gamma=1.29\pm0.05$ and $\Gamma=1.64\pm0.05$ respectively, compared to the 2007 spectrum ($\Gamma=2.25\pm0.03$). This simple single power-law model provided a statistically very poor fit to the data, with $\chi^{2}/{\nu}=2125/1318$. Furthermore, it is thought that the underlying continuum in \pds is intrinsically steep ($\Gamma\gtrsim2$), as evident from the 2007 observation, when the continuum was observed through little or no obscuration (R09), as well as from the \nustar observations, with typically $\Gamma\sim2.4$ measured above $10$ keV (e.g., N15). It follows that such a drastic hardening of the 2013 \suzaku spectra may be unphysical if simply attributed to the shape of the intrinsic continuum. On the other hand, what is seen is a series of spectra that may be affected by either complex absorption or reprocessed (reflected/scattered) emission, or the contribution of both. In addition to the harder shape of the three 2013 sequences, a deeper absorption trough is observed at $\sim9$ keV (in the quasar rest-frame) which apparently strengthens as the observation progresses from 2013a to 2013c. In Table~\ref{tab:optxagnf} we list the iron K emission and absorption line properties as parameterised by Gaussian profiles, whose variability will be discussed later in section~\ref{The Iron K Band: Emission and Absorption Profiles}. 
\\ 
Generally speaking, it has been suggested that all AGN characterised by a low flux spectrum may be reflection-dominated \citep[e.g.,][]{Fabian12b,Gallo13}. Nonetheless, there are some clear differences in the \pds spectra, such as the non detection of a \textquotedblleft Compton hump" above $10$ keV (see Fig.~\ref{fig:pds_xmm_nu_suz_good} and N15), and the lack of a narrow $6.4$ keV iron K$\alpha$ line (see Fig.~\ref{pds456_5seq_time_fek}), which even in the 2013 low flux state has a low upper limit on the equivalent width (EW) of $<28$ eV (2013a), $<31$ eV (2013b) and $<37$ eV (2013c). This rules out the presence of a distant Compton-thick reprocessor. Neither is a prominent red wing to any broad iron K emission evident, unlike what claimed in the spectra of other AGN \citep{Fabian02}. This suggests that \pds is probably not dominated by strong Compton reflection components, from either the disc or distant matter. This alternative physical interpretation and the comparison between reflection and absorption based models will be investigated in detail in a parallel paper (Costa et al.:in preparation).

\subsection{Modelling The broadband SED}
\label{sebsec:Modelling The broadband SED}

The spectral variability that characterises \pds is conspicuous, in particular over a six month/one year timescale, as shown in Fig.~\ref{fig:pds_xmm_nu_suz_good}. In order to gain a better understanding of this pronounced spectral variability, we first need to characterize the broadband intrinsic continuum of PDS 456. We therefore tested whether the broadband SED of PDS 456 could be described by a multi-temperature Comptonised accretion disc model, using the \texttt{optxagnf} model \citep{Done12} in XSPEC. This model is characterised by three separate components, which are self-consistently powered by dissipation in the accretion flow: (i) the thermal emission from the outer accretion disc in the optical/UV; (ii) the up-scattering of the UV disc photons into a soft X-ray excess from a warm disc atmosphere; (iii) a high temperature Comptonisation from the corona (i.e. the standard hard X-ray power-law continuum). The parameter r$_{\rm cor}$ is the coronal size that acts as a transitional radius from the colour temperature corrected blackbody emission (produced from the outer disc) to a hard power law produced through Compton up-scattering. The parameter \texttt{F$_{\rm pl}$} gives the fraction of the energy released in the power-law component. The parameters $kT$ and $\tau$ are the electron temperature and the optical depth of the soft Comptonisation component, possibly originating from the warm disc atmosphere and seen as the soft excess, while $\log({\rm L}/{\rm L_{Edd}})$ is the Eddington ratio of the AGN (see \citealt{Done12} for more details).\footnote{In this work, for simplicity, the spin parameter \texttt{a$^{\star}$} was kept fixed to zero in all the \texttt{optxagnf} fits.}

\begin{figure}
\begin{center}
\includegraphics[width=10cm]{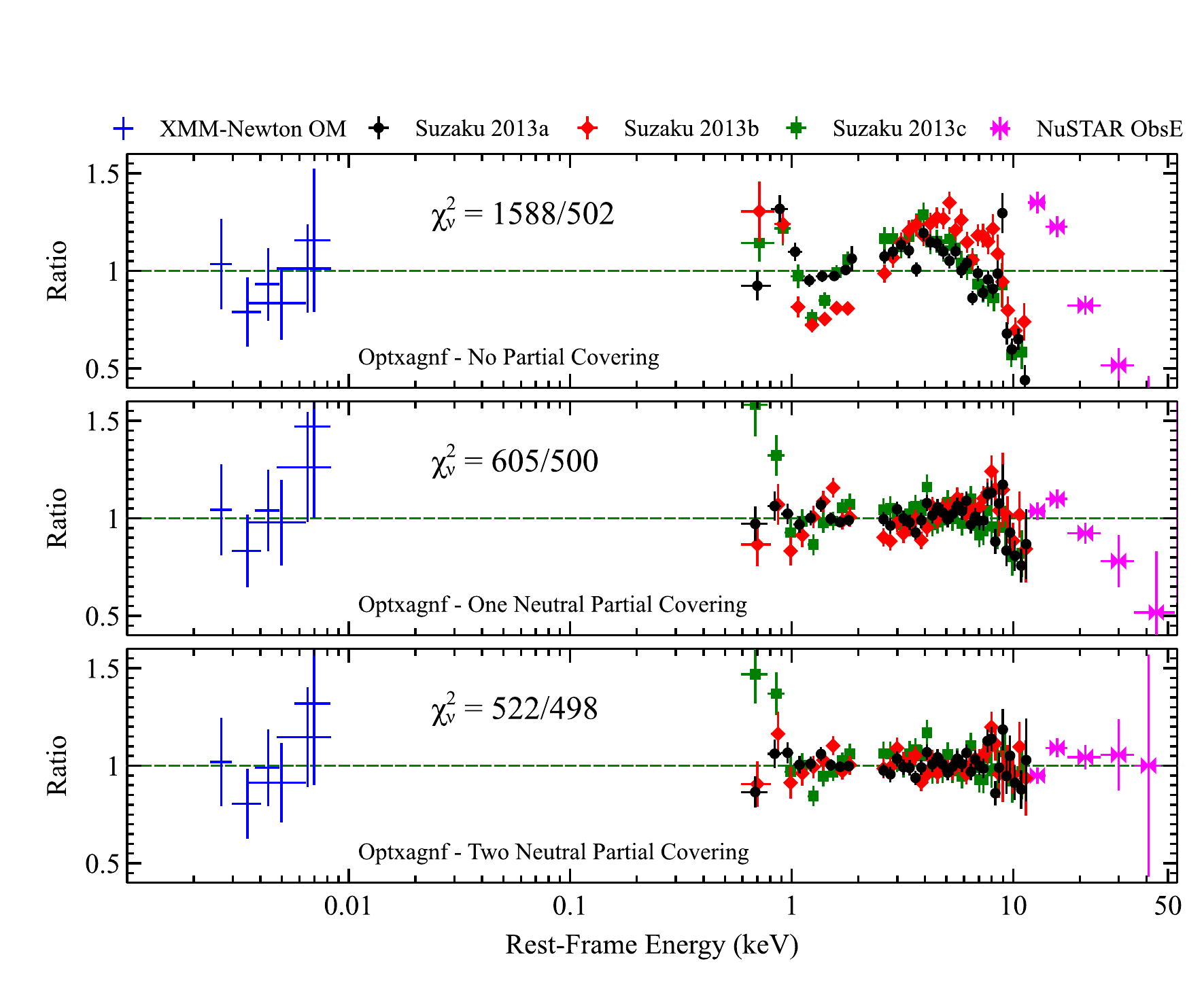}	
\caption{Residuals for the different \texttt{optxagnf} model fits over the $1$ eV - $50$ keV energy range, plotted as data/model ratios. The spectra correspond to the \xmm OM and \nustar from Obs E in 2014 (blue and magenta respectively), while also plotted are the \suzaku 2013a (black), 2013b (red), 2013c (green) sequences. Top panel: SED fits with \texttt{optxagnf} and no partial covering absorption. Middle panel: Same as above but with the addition of one partial covering layer. Bottom panel: Same as above but with two partial covering layers. For extra clarity, the spectra have been re-binned.} 
\label{pds_optxagn_ratio_zpcs}
\end{center}
\end{figure}

The form of the optical/UV to hard X-ray SED has been defined when the X-ray observations are in the 2013 \suzaku low flux state. To this end, we also make use of the optical/UV photometry provided by the \xmm Optical Monitor, noting that, although the optical/UV fluxes are not simultaneous with the \suzaku observation six months earlier, there appears to be little variability in this band from both the \xmm observations (to within $5\%$) and the archival \textit{Swift} observations over a period of $\sim6$ months (A. Lobban; private communication). In order to define the hard X-ray spectral shape above $10$ keV, we consider only the lowest flux (Obs E) \nustar spectrum, as it appears closest in flux to the \suzaku 2013a observations at $10$ keV and is also consistent with the upper limit obtained for the HXD/PIN (see Tables~\ref{tab:suzaku}, \ref{tab:xmm/nustar} and Fig~.\ref{fig:pds_xmm_nu_suz_good}). The datasets were fitted simultaneously but allowing for a cross-normalization factor between the \suzaku 2013 and \nustar (Obs E) to account for any absolute differences in hard X-ray flux, which is found to be very close to $1.0$ (i.e. $0.98\pm0.08$).
\\
When fitting the SED with \texttt{optxagnf}, we found that some absorption is required to account for the spectral curvature above $1$ keV, as shown in Fig.~\ref{pds_optxagn_ratio_zpcs} (top panel). By adding one layer of neutral partial covering (\texttt{zpcfabs}) of column density $\log(N_{\rm H}/{\rm cm^{-2}})=22.47_{-0.04}^{+0.02}$ and covering fraction $f_{cov}=0.97\pm0.01$ (tied between the three sequences), the fit improved significantly by $\Delta\chi^{2}/\Delta\nu=983/2$. However, some curvature is still present in the residuals above $\sim3$ keV, and the steepness of the spectral slope above $10$ keV, observed with the \nustar data, is not reproduced (Fig.~\ref{pds_optxagn_ratio_zpcs}, middle panel). By adding a second layer of neutral partial covering, the fit improved by a further $\Delta\chi^{2}/\Delta\nu=83/2$, yielding $\chi^{2}/{\nu}=522/498$. In this latter case, a lower column zone (\texttt{zpcfabs}$_{\rm low}$) of column density $\log(N_{\rm H,low}/{\rm cm^{-2}})=22.4\pm0.1$ and covering fraction $f_{cov,\rm low}=0.88_{-0.03}^{+0.02}$ is required to account for the absorption present in the soft X-ray band. A higher column zone (\texttt{zpcfabs}$_{\rm high}$), with column density of $\log(N_{\rm H,high}/{\rm cm^{-2}})=23.1\pm0.1$ and covering fraction $f_{cov,\rm high}=0.71_{-0.09}^{+0.10}$, parameterises the spectral curvature above $2$ keV; thus it is the combination of the two zones that reproduces the complex overall spectral curvature below $10$ keV. 
\\
We initially kept the partial covering column density and covering fractions tied between the three \suzaku sequences; in order to allow for small changes we let the \texttt{optxagnf} parameters $\tau$ and F$_{\rm pl}$ free to vary. The coronal size was also tied between the sequences, and it is found to be $r_{\rm cor}=41_{-19}^{+58}$ $R_{\rm g}$. Due to the degeneracy between the optical depth and the temperature of the warm electrons responsible for the soft excess, we kept the latter tied between the observations, yielding $kT =468_{-107}^{+156}$ eV. The Eddington ratio $\log({\rm L}/{\rm L_{Edd}})=-0.08$ implies that \pds radiates close to its Eddington luminosity ($\sim 80\%$ of $L_{\rm Edd}$); this is consistent with the expectations for PDS\,456, given its black hole mass ($M_{\rm BH}\sim10^{9}\Msun$) and bolometric luminosity ($L_{\rm bol}\sim10^{47}$ erg s$^{-1}$). When allowing the covering fractions to vary between the three sequences, they are typically found to be within $\sim10\%$ of each other, and the statistical improvement is not too drastic ($\Delta\chi^{2}/\Delta\nu=13/2$), at least considering these time-averaged spectra. We note that the value of $r_{\rm cor}$ above is somewhat larger than usually expected \citep[e.g.,][]{Risaliti09b,Reis13}, but given our assumptions (for instance, zero black hole spin) this is regarded as a rough estimate only. If we fix $r_{\rm cor}=10$ $R_{\rm g}$ (see section~\ref{sec:Discussion}) we recover a fit equivalent to the case with \texttt{a$^{\star}=0$} above  for \texttt{a$^{\star}\sim0.5$}. In Section 5 we return to consider the spectral variability in the time-sliced spectra, where more pronounced variability is present on shorter timescales.

\section{The Fe K band Modelling}

Having parameterised the continuum form in the previous section, here we focus on the analysis of the properties of the iron K emission and absorption profiles and on their variability over the 2013 \suzaku observations.

\subsection{The Iron K Band: Emission and Absorption Profiles}
\label{The Iron K Band: Emission and Absorption Profiles} 

When the absorption feature, in the \texttt{optxagnf} baseline model, was fitted with a Gaussian profile (\texttt{zgauss$_{abs}$}), the centroid energy at $\sim8.6-8.8$ keV indicates a large degree of blueshift when compared to the expected lab-frame energies of the \textit{1s-2p} lines of He- or H-like Fe at $6.7$ and $6.97$ keV. The absorption line EW increases by a factor of $\sim5$ throughout the observation, from EW $=-126\pm63$ eV in the 2013a sequence (improving the fit by $\Delta \chi^2/\Delta\nu=15/2$) to EW $=-353\pm82$ eV in 2013b ($\Delta \chi^2/\Delta\nu=61/2$), and EW $=-530\pm72$ eV in 2013c (improving considerably the fit by $\Delta \chi^2/\Delta\nu=141/2$; see Table~\ref{tab:optxagnf}). 
\\
A cosmologically redshifted ($z=0.184$) Gaussian component (\texttt{zgauss$_{em}$}) parameterises the ionised emission profile at $6.9\pm0.1$ keV (in the quasar rest frame), likely corresponding to the Fe \textsc{xxvi} Ly$\alpha$ resonance line, although we found that the emission profile is generally difficult to constrain in these 2013 data sets. So the nature of the iron K emission might be associated with the P-Cygni-like feature (e.g., see the Fe K profile in 2013b,c from Fig.~\ref{pds456_5seq_time_fek}) arising from the disc wind and resolved in the simultaneous \xmm and \nustar observations carried out in late 2013/early 2014 (N15). Similar to the fit adopted in N15, the widths of the emission and the absorption Fe K profiles were tied together with a common velocity broadening of $\sigma=264_{-56}^{+74}$ eV at the energy of the absorption line, or $\sigma=212$ eV at $7$ keV (this corresponds to a maximum velocity dispersion of $\sigma_{\rm v}\sim9100~ \rm km~s^{-1}$ or FWHM $\sim21400~ \rm km~s^{-1}$). This model provided a statistically very good fit ($\chi_{\nu}^{2}=522/498$). Unsurprisingly, if the Fe K absorption feature is not accounted for the model will produce a very poor fit, i.e. $\chi_{\nu}^{2}=717/502$.

\subsubsection{X-ray Background}

\begin{figure}
\begin{center}
\includegraphics[width=8cm]{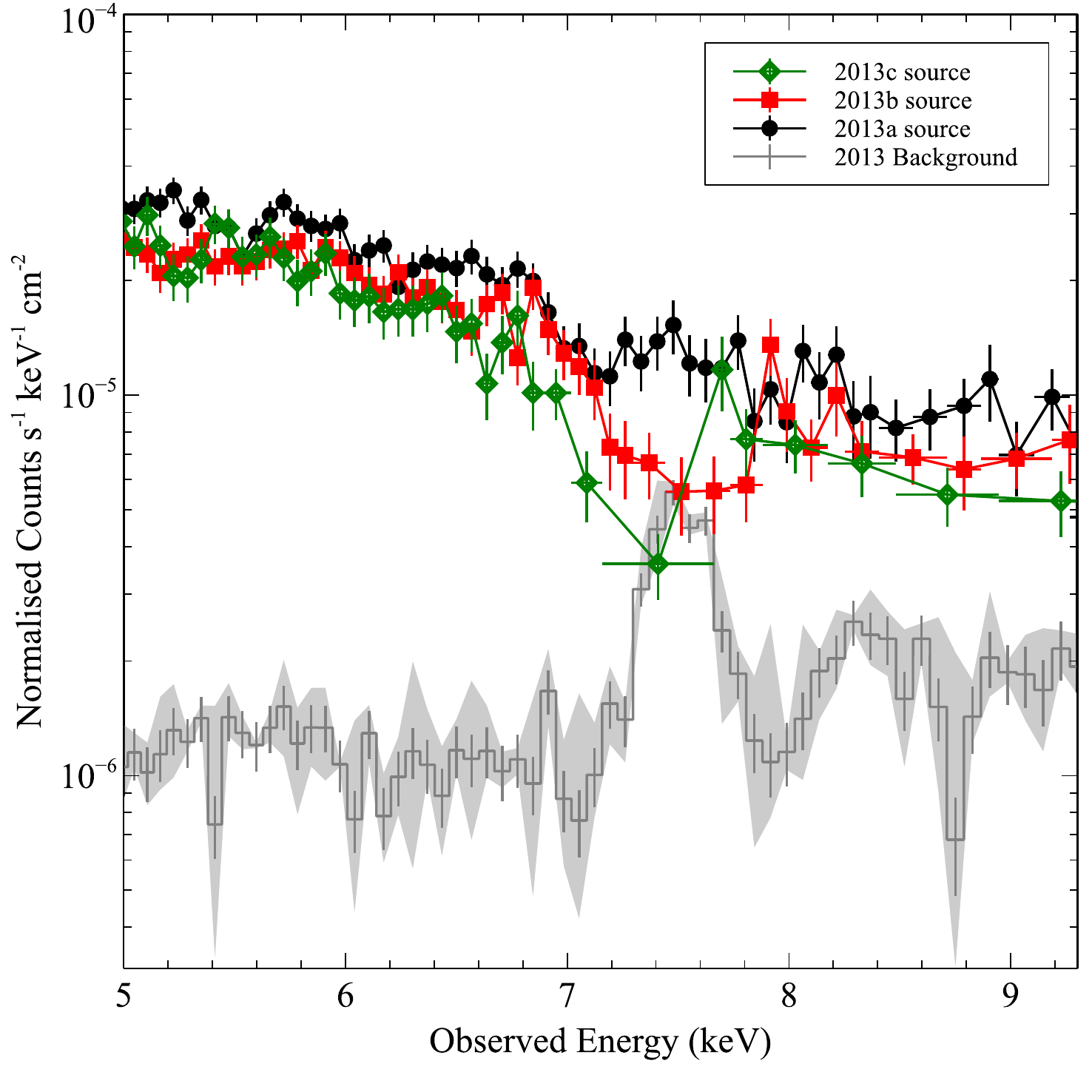}	
\caption{2013 \suzaku XIS03 background subtracted source spectra versus the averaged background spectrum. The shaded area indicates the degree of the background fluctuation across the observation. A prominent background line, corresponding to Ni K$\alpha$, arises at $\sim7.5$ keV; this was also detected by R09 in the 2007 \pds spectrum (see their Fig. 3). Unfortunately, in the 2013 spectra the centroid energy of the Ni K$\alpha$ corresponds to that of the Fe K absorption feature, as the latter is shifted to lower energies compared to the 2007 observation. It is important to note the amplitude of the background variability of $\pm10\%$ is too small to influence the strong increase in strength of the Fe K absorption profile from the 2013a through to the 2013c observation. The source spectra are grouped to a 4$\sigma$ significance for each energy bin.}
\label{bkg}
\end{center}
\end{figure}

Fig.~\ref{bkg} shows the comparison (focused on the Fe K band) between the net source spectra and the averaged background spectrum, where the shaded area indicates its maximal fluctuation during the observation. A strong background emission line, corresponding to Ni K$\alpha$, arises at $\sim7.5$ keV, which coincides with the observed frame energy of the Fe K absorption profile during the 2013 \suzaku observations. Thus it is possible that, if the background is incorrectly subtracted, it could contribute towards the absorption line feature in the source spectrum. The absorption feature is anyway variable by a factor of $\sim4-5$, whilst the background is almost constant ($\pm10\%$) across all of the 2013 observations. To be conservative, we fitted the background spectrum of 2013c (where the Fe K absorption feature is strongest) between $5-9$ keV with a simple power law and a Gaussian emission profile, centred at $7.49\pm0.03$ keV. We find that by comparing the normalisations of both the Ni K$\alpha$ emission ($\sim1.2\pm0.3\times10^{-6}$ photons cm$^{-2}$ s$^{-1}$) and the Fe K absorption ($\sim-9.6\pm1.3\times10^{-6}$ photons cm$^{-2}$ s$^{-1}$) lines, the former could only contribute to $\sim12\pm3\%$ of the latter. Consequently, there is no possibility that the background feature can account for both the observed strength and variability of the absorption line, as we would expect similar variability amplitudes from the absorption line and the Ni background line, a trend that we clearly do not observe. Moreover, in the 2007 \suzaku observation (R09) the Fe K absorption lines were observed at higher energies (at $7.67$ keV and $8.15$ keV respectively) and thus were clearly separated from the Ni K$\alpha$ feature at $7.45$ keV (see Fig. 3 in R09). The level of the background relative to the source was much lower in those observations, as the source was much brighter. During the \xmm observations of \pds (N15), the overall level of the background is substantially lower ($<1\%$ compared to the source spectrum), due to the smaller extraction region used. Thus for either \suzaku or \xmm, the background subtraction likely makes a negligible contribution towards the absorption line.

\subsection{Input SED for \textsc{xstar}}

\begin{figure}
\begin{center}
\includegraphics[width=8cm]{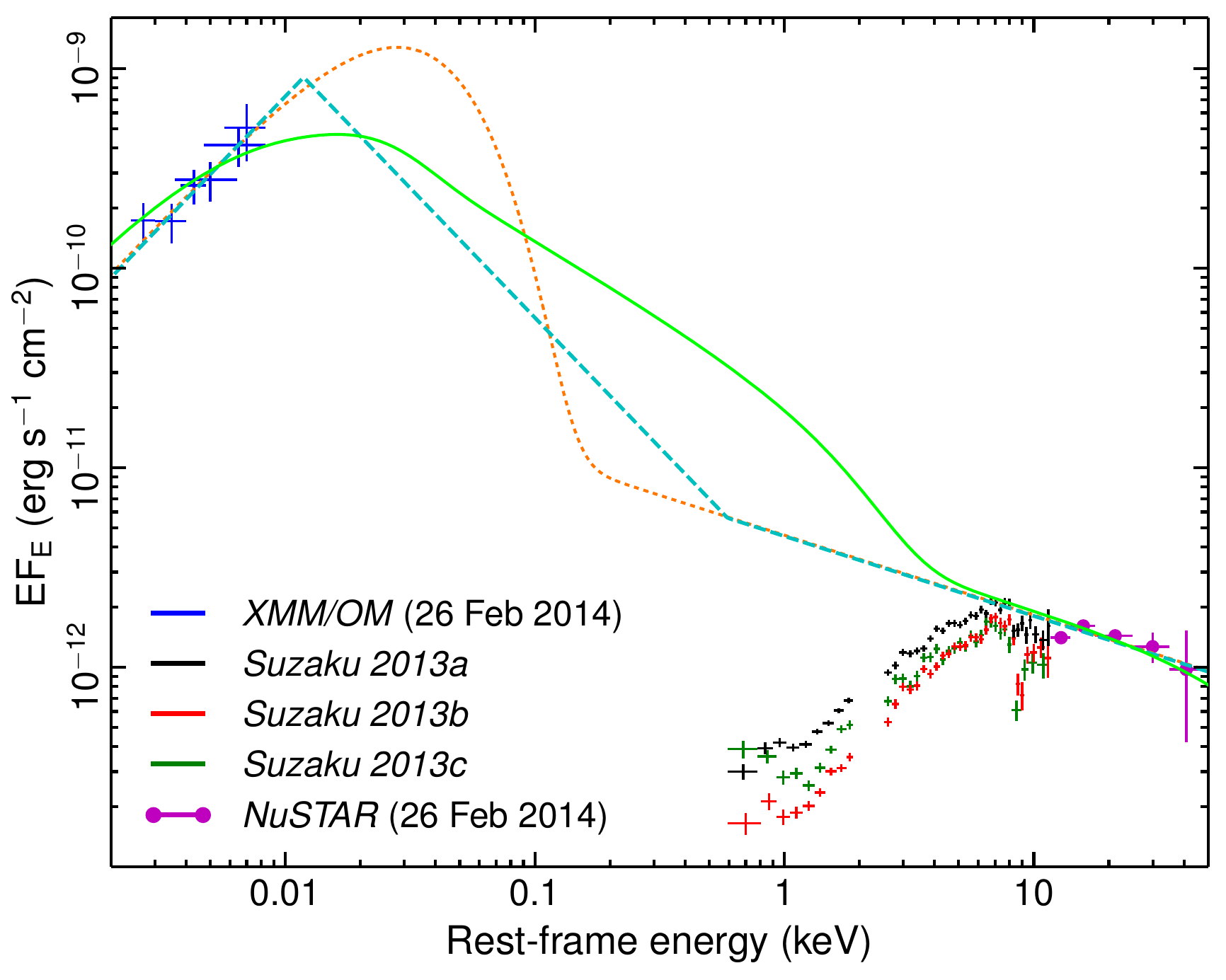}	
\caption{Optical to hard X-ray Spectral Energy Distribution (SED) of \pds obtained by combining the OM and \nustar (blue and magenta respectively) from the last \textit{XMM-Newton/NuSTAR} observation in 2014 and the 2013a,b and c sequences from the \suzaku campaign (black, red and green respectively). The SED was first fitted with the \texttt{optxagnf} model (green line), and then approximated with a double broken power law (dashed cyan line) with break energies of 10 eV and 500 eV as an input for \textsc{xstar}. For comparison only, the dotted orange curve shows a disc blackbody component with temperature of 10 eV. Note that the models are corrected for absorption (either Galactic or local to the source) whereas the spectra are not.}
\label{fig:pds456_sed_om_suz}
\end{center}
\end{figure}

Having parameterised the iron K absorption with a simple Gaussian models, we proceed to model the iron K absorption in \pds with multiplicative grids of photoionised spectra. In order to do this we first need to characterize the input photoionising continuum. Self consistent \textsc{xstar} \citep[v2.21bn13,][]{BautistaKallman01} grids were generated using the UV to hard X-ray SED of PDS 456. Previously, G14 adopted an \xstar absorption grid with velocity broadening $\sigma_{\rm turb}=5000$ km s$^{-1}$, using a single power law of $\Gamma=2.4$ for the 1-1000 Rydberg continuum. Here, following N15, we estimated the SED of \pds by using the simultaneous \xmm and \nustar data (including six photometric bands of the Optical Monitor, see section~\ref{sebsec:Modelling The broadband SED}). We found that the SED can be approximated with a phenomenological double broken power law (see Fig.~\ref{fig:pds456_sed_om_suz}). This yielded $\Gamma\sim0.7$ in the optical/UV up to an assumed break energy of $10$ eV, $\Gamma\sim3.3$ between the optical/UV and the soft X-ray band up to the second break energy fixed at $0.5$ keV, and $\Gamma=2.4$ beyond that in the X-ray band (see Fig.~\ref{fig:pds456_sed_om_suz}). By adopting this model, we estimated a total $1-1000$ Rydberg ionising luminosity of $L_{\rm ion}\sim5\times10^{46}$ erg s$^{-1}$, which is also broadly consistent with the estimate derived earlier for the \texttt{optxagnf} model (Fig.~\ref{fig:pds456_sed_om_suz}).
\\
\noindent Subsequently, we adopted this SED as the input continuum for our custom \xstar grids. We investigated the effects of different velocity broadening by generating various \xstar grids for a range of $\sigma_{\rm turb}$ values in order to provide an accurate description of the width of the Fe K absorption profile. We found that grids with lower $\sigma_{\rm turb}$ (e.g., $\sigma_{\rm turb}=1000-3000$ km s$^{-1}$) result in a smaller EW of the absorption lines compared to the data, as the absorption profile saturated too quickly at a lower column density. An adequate fit to the Fe K absorption profile was achieved when we adopted a grid with $\sigma_{\rm turb}=10000$ km s$^{-1}$; this is consistent with the velocity width measured earlier from a simple Gaussian profile. Thus we used the $\sigma_{\rm turb}=10000$ km s$^{-1}$ grid, with the input SED described above, in all the subsequent fits.

\subsection{Photoionisation Modelling of the Fe K Absorption}

\label{subsec:Photoionisation Modelling of the Fe K Absorption}

\begin{figure*}
\begin{center}	

\includegraphics[scale=0.6]{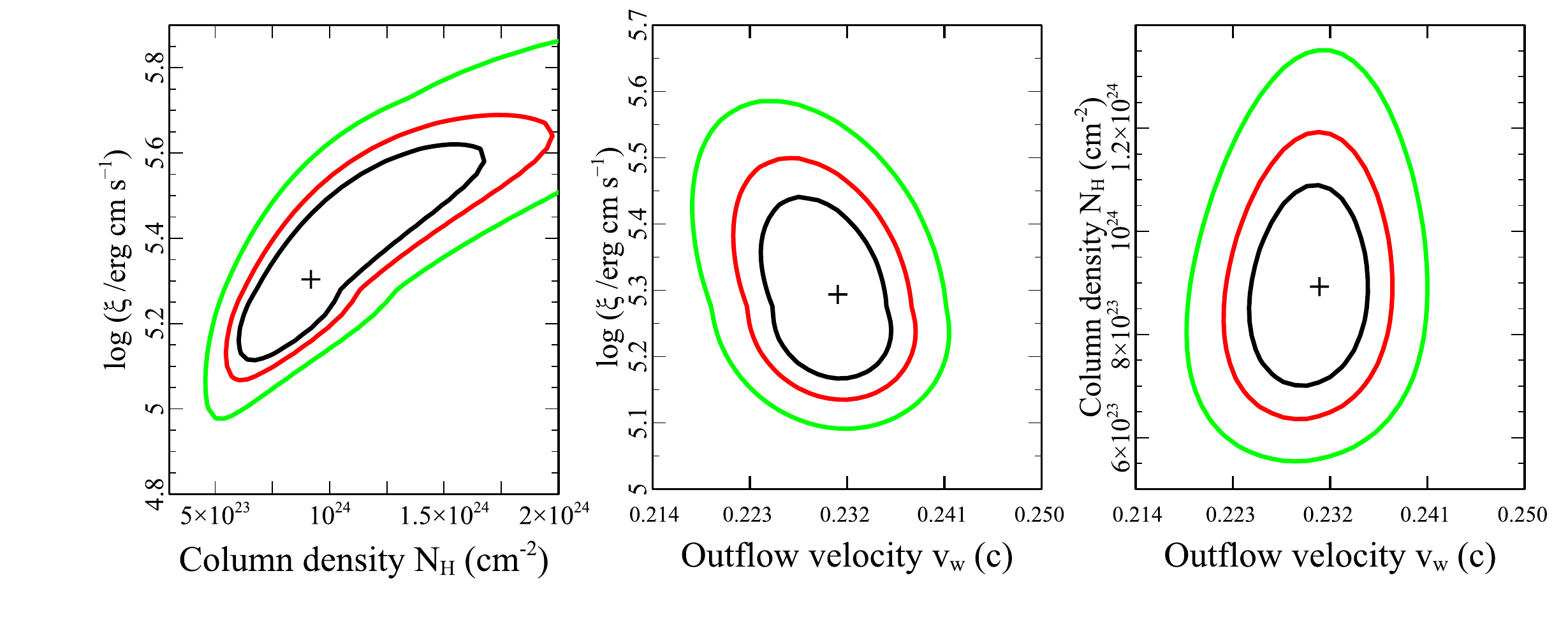}	

\end{center}
\caption{Two dimensional contour plots from the \xstar parameters of the highly ionised wind in the 2013c segment. The solid black, red and green lines correspond to 68 $\%$, 90 $\%$ and 99 $\%$ confidence levels for two interesting parameters respectively. The elongated contour between the ionization against the column density parameters in the left panel indicates a possible degeneracy. On the other hand, the other two contour plots (centre and right panel) show no apparent degeneracies between either of the former parameters and the outflow velocity.}
\label{fig:pds456_2013c_2D_nh_xi_z}
\end{figure*}

Having generated a suitable \xstar grid, we used it to parameterise the Fe K absorption line. We first applied \xstar to the 2013c spectrum, where the iron K absorption profile is stronger, in order to investigate any physical degeneracy that may arise in the model by allowing the column density and the ionisation parameter\footnote{The ionisation parameter is defined as $\xi=L_{\rm ion}/nR^{2}$, where $L_{\rm ion}$ is the $1-1000$ Rydberg luminosity, $n$ is the electron density of the gas and $R$ is the distance of the ionising source from the absorbing clouds.} to vary along with the absorber's velocity. In Fig.~\ref{fig:pds456_2013c_2D_nh_xi_z} (left panel), the elongated shape of the contours does indeed imply that there is some degeneracy between $N_{\rm H}$ and $\log(\xi)$. This behaviour can be attributed to the fact that most of the iron is in the He- and H-like state. Thus a gradual increase of $\log(\xi)$ would not significantly affect the line centroid energies, but it would instead increase the proportion of fully ionised iron, requiring an apparent increase in $N_{\rm H}$ to compensate for it. On the other hand, the other two contour plots (centre and right panel) show no apparent degeneracies between either $N_{\rm H}$ or $\log(\xi)$ and the outflow velocity. On this basis, and given the degeneracies between $N_{\rm H}$ and $\log(\xi)$, the variability of the iron K absorption feature has been parameterised by either (i) allowing only the ionization to vary over the course of the observation with constant $N_{\rm H}$, or (ii) allowing only the column density to vary with constant $\log(\xi)$. In both scenarios we kept the outflow velocity constant at (i) $v_{\rm w}=0.25\pm0.01c$ and (ii) $v_{\rm w}=0.24\pm0.01$c between the sequences for each case above.\footnote{The outflow velocity shows some variability during the observation in the range $v_{\rm w}\sim0.23c$ to $0.27c$ (see section \ref{Time-sliced spectra}). However, assuming a constant velocity is equivalent on statistical grounds.} In case (i) we find that the ionisation of the line varies as $\log(\xi/{\rm erg~cm~s^{-1}})=6.6_{-0.4}^{+0.9}$, $\log(\xi/{\rm erg~cm~s^{-1}})=6.0_{-0.3}^{+0.2}$ and $\log(\xi/{\rm erg~cm~s^{-1}})=5.4_{-0.3}^{+0.2}$ in 2013a, 2013b and 2013c, respectively, by keeping the column density tied at $\log(N_{\rm H}/{\rm cm^{−2}})=24.0_{-0.3}^{+0.1 }$. In case (ii) the column density of the absorption line varies as $\log (N_{\rm H}/\rm cm^{-2})=23.1\pm0.2$, $\log (N_{\rm H}/\rm cm^{-2})=23.6_{-0.1}^{+0.3}$ and $\log (N_{\rm H}/\rm cm^{-2})=24.0\pm0.2$ in 2013a, 2013b and 2013c, respectively, for a constant ionisation of $\log$($\xi$/erg cm s$^{-1}$) $=5.3_{-0.1}^{+0.3}$. Further implications regarding the Fe K absorption variability will be discussed more in detail in section~\ref{subsec:The Iron K Short-Term Absorption Variability} below, where the variations are investigated on a shorter time-scale across the \suzaku 2013 campaign by means of time-sliced spectroscopy.

\section{Time-sliced spectra}
\label{Time-sliced spectra}

\begin{figure*}
\begin{center}	
\includegraphics[scale=0.50]{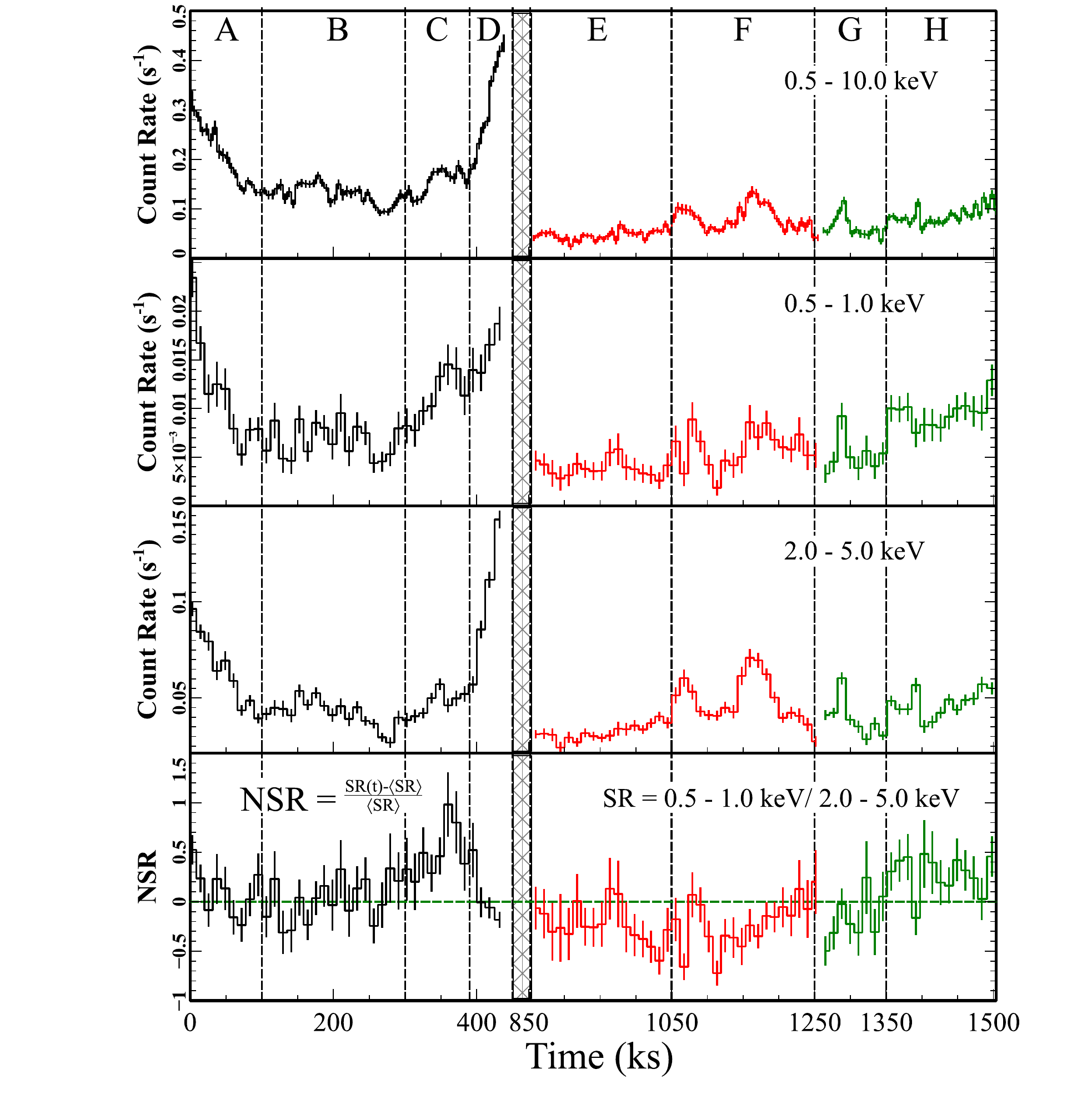}	
\caption{Light curves and normalised softness ratio ($0.5-1/2-5$ keV) for sequences 2013a (black), 2013b (red) and 2013c (green). The dashed vertical lines identify the boundaries of the eight slices.} 
\vspace{-2.5mm}
\begin{tablenotes} 
 	\item[a] Panel 1: XIS-FI (X-ray Imaging Spectrometer-Front Illuminated) $0.5-10.0$ keV light curve of the overall \suzaku observation. Note the strong flare between $400-450$ ks in segment D.	
	\item[b] Panel 2: $0.5-1$ keV soft band light curve.
	\item[c] Panel 3: $2-5$ keV hard band light curve. 	
 	\item[d] Panel 4: Fractional change in the softness ratio (see the text for the definition). Positive (negative) values correspond to the softening (hardening) of the source with respect to the average spectral state. The bin size of the light curves are 5760 s, corresponding to one satellite orbit.
\end{tablenotes}
\label{fig:pds456_2013_3lc_1sr}
\end{center} 
\end{figure*}

\begin{figure*}
\centering	
\includegraphics[scale=0.45]{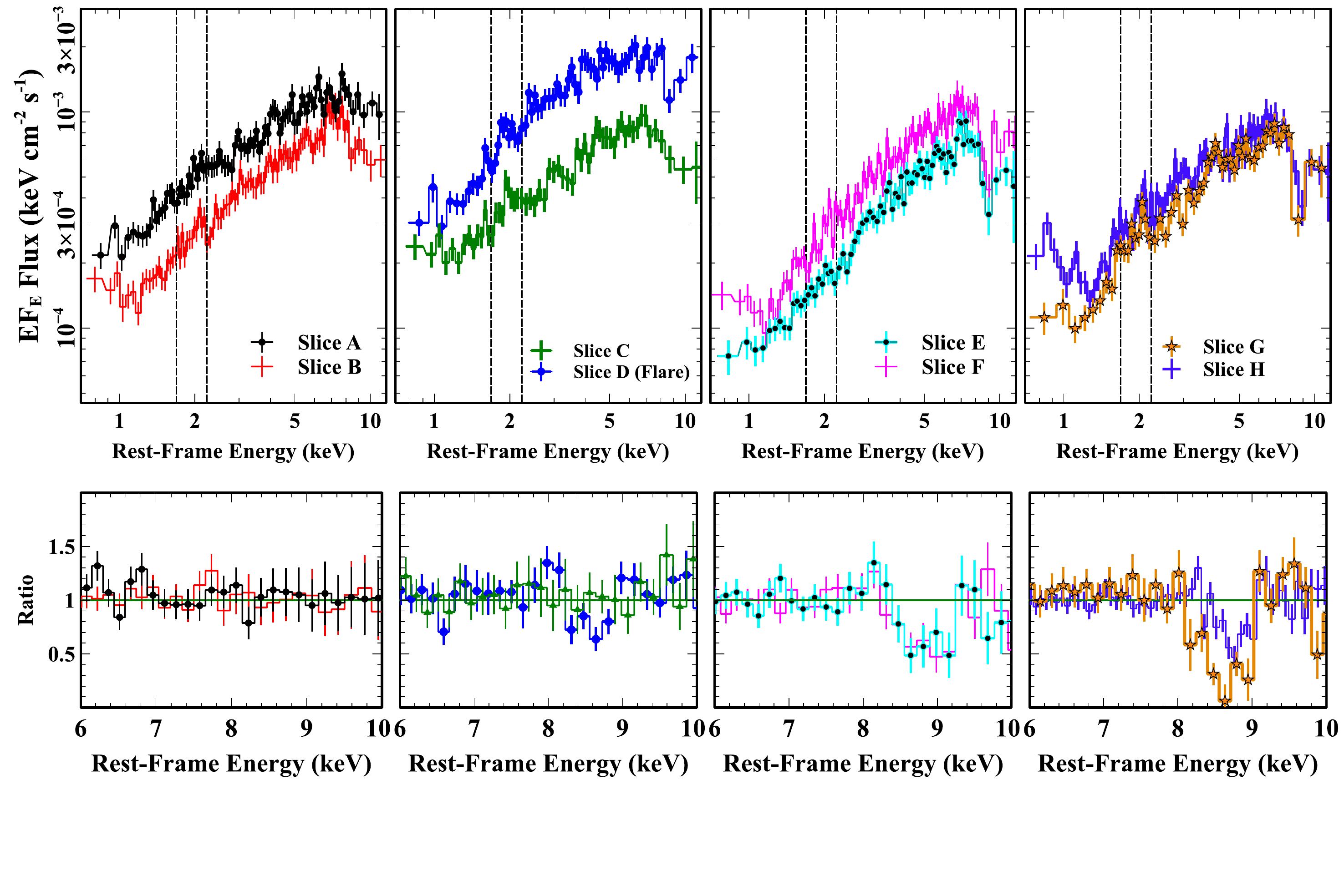}	
\vspace{-10mm}
\caption{Top panels: Fluxed spectra of the slices A (black), B (red), C (green), D (blue), E (cyan), F (magenta), G (orange) and H (violet). What is noticeable is the absence of a clear Fe K absorption feature in the first three spectra, and its possible onset in slice D (corresponding to the flare). The dotted rectangular areas represent the XIS Si edge calibration uncertainty range between 1.7-2.1 keV, ignored during fitting. The depth of the Fe K absorption feature ($\sim7.4$ keV in the observed frame, $\sim8.7$ keV in the quasar rest frame) gradually increases after the flare between slice E and slice G, when the absorption trough becomes very prominent. The fluxed spectra have been plotted against a simple power-law with $\Gamma=2$ with minimum 50 cts per bin. Bottom Panels: Ratio spectra of the the slices in the Fe K band obtained as in Fig.~\ref{pds456_5seq_time_fek}.}
\label{fig:pds456_eeuf_gamma2_8slice}
\end{figure*}

Fig.~\ref{fig:pds456_2013_3lc_1sr} shows the overall light curves of the 2013 \suzaku campaign, strongly indicating variability of the X-ray flux in \pds on short time scales. A prominent flare is detected, with the flux increasing by a factor of $4$ between $400-450$ ks in sequence 2013a, followed by smaller flares towards the second half of sequence 2013b. Guided by the visual properties of the overall light curve and softness ratio (between the $0.5-1$ and $2-5$ keV bands), the spectra were divided into a total of eight slices (see panel 1 in Fig.~\ref{fig:pds456_2013_3lc_1sr}); this was done by taking into consideration the width of each slice and the number of counts in it. Note that for PDS\,456, with $M_{\rm BH}\sim10^{9}\Msun$, a variability timescale of $\sim100$ ks corresponds to a light-crossing distance of $\sim20$ R${\rm _g}$. A slice of at least $100$ ks in duration was usually required for the 2013 spectra whilst, on the other hand, we had enough counts to isolate the much brighter flare between $400-450$ ks into the observation. The plots of the fluxed $\nu F_{\nu}$ spectra and the Fe K band data to model ratios (obtained as in Fig.~\ref{pds456_5seq_time_fek}) of the eight slices are shown in the top and bottom panels of Fig.~\ref{fig:pds456_eeuf_gamma2_8slice}. The first four spectra (A - D), correspond to 2013a sequence, tracing the decline of an initial flare (A) followed by a quiescent period (B) together with the initial onset of the large flare (C) and the subsequent flare itself (D). In the remaining four spectra (E - H), corresponding to the 2013b and 2013c sequences, the Fe K absorption feature becomes very noticeable (E), progressing in strength (F) to reach maximum depth in slice G and tentatively recovering in slice H. The timing periods for each of the slices are noted in Table~\ref{tab:fek_gaussian}.
\\
In section~\ref{sebsec:Modelling The broadband SED} we used the \texttt{optxagnf} model to account for the optical/UV to hard X-ray SED of PDS\,456. However, in order to describe the lower S/N time-sliced spectra over the $0.6-10$ keV band, we used a simpler two-component model to provide a more convenient parameterisation of the intrinsic continuum. We therefore parameterised the continuum with a phenomenological baseline model (hereafter MODEL I) of the form:

\begin{equation}
\begin{split}
\texttt{Tbabs}\times[\texttt{zpcfabs}_{\rm low}\times\texttt{zpcfabs}_{\rm high}\times(\texttt{po}+\texttt{bbody})\\+\texttt{zgauss$_{em}$}  +\texttt{zgauss$_{abs}$}],  
\end{split}
\end{equation}

\noindent where \texttt{Tbabs} accounts for the Galactic absorption using the cross sections and ISM abundances of \citet{Wilms00}. The Gaussian component \texttt{zgauss$_{em}$} parameterises the ionised emission profile at $6.9\pm0.1$ keV (in the quasar rest frame). We note that this two component (power law plus blackbody) continuum model, when applied to the parent sequences, is fully consistent with the earlier \texttt{optxagnf} findings. Again, the model requires two neutral partially covering absorbers, with column densities of $\log(N_{\rm H,low}/{\rm cm^{-2}})=22.3\pm0.1$ and $\log(N_{\rm H,high}/{\rm cm^{-2}})=23.2\pm0.1$. The soft excess, now parameterised with a blackbody (\texttt{bbody}), yields a temperature of kT $\sim82_{-20}^{+21}$ eV, while the properties the Fe K emission and absorption profiles, as expected, are not modified compared to Table~\ref{tab:optxagnf}.

\subsection{The Iron K Short-Term Absorption Variability}
\label{subsec:The Iron K Short-Term Absorption Variability}

Here we investigate the variability of the iron K absorption when applied to the short spectral slices. To provide an acceptable fit to the baseline continuum, we applied MODEL I introduced above over the $0.6-10$ keV band. The high and low column covering fractions ($f_{\rm cov,low}$ and $f_{\rm cov,high}$) are allowed to vary between the eight slices, together with the power-law and blackbody normalisations. The high and low column densities of the partial covering absorbers were tied between the slices.

\subsubsection{Gaussian Modelling}

To characterise the behaviour of the Fe K absorption line across all the eight slices, this has been initially parameterised using a simple Gaussian profile. Statistically speaking, we found that the Fe K absorption is not significant (at $>99\%$ confidence level) in the first three slices (A - C). As the observation progresses the absorption feature becomes significant, through slices E ($\Delta\chi^{2}/{\Delta\nu}=24/2$, equivalent to $\sim5\sigma$ confidence level) and F ($\Delta\chi^{2}/{\Delta\nu}=40/2$ $\sim6\sigma$), reaching its maximum depth during slice G ($\Delta\chi^{2}/{\Delta\nu}=77/2$~$>8\sigma$), with EW $=-537^{+121}_{-131}$ eV, when the count rate becomes effectively null at the centroid of the absorption line ($E=8.63\pm0.11$ keV; see Fig.~\ref{fig:pds456_eeuf_gamma2_8slice}). Assuming a common velocity width between all the slices as in section~\ref{The Iron K Band: Emission and Absorption Profiles}, we obtained $\sigma=282_{-56}^{+69}$ eV, thus corresponding to a velocity dispersion of $\sim9900$ km s$^{-1}$. Model I provided a statistically very good fit to the slices, with $\chi_{\nu}^{2}=1554/1681$. The parameter details of the Gaussian absorption line, for each of the eight slices, are listed in Table~\ref{tab:fek_gaussian} and plotted in Fig.~\ref{fig:pds456_3panels_nh_xi_wind}(a).

\begin{table*}
\begin{tabular}{ccccc}



\hline
Slice&Time&$E$ (keV)$^a$&EW (eV)$^b$&($\Delta \chi^{2}/\Delta\nu$)$^c$\\
\hline

A&0-100&$8.55^t$&$<105$&--\\

B&100-300&$8.55^t$&$<156$&--\\

C&300-400&$8.55^t$&$<145$&--\\

D&400-450&$8.55\pm0.15$&$254_{-89}^{+116}$&$15/2$\\

E&850-1050&$8.84\pm0.14$&$341_{-121}^{+112}$&$24/2$\\

F&1050-1250&$8.85_{-0.11}^{+0.12}$&$350_{-90}^{+100}$&$40/2$\\

G&1250-1350&$8.63\pm0.11$&$537^{+121}_{-131}$&$77/2$\\

H&1350-1510&$8.69\pm0.11$&$396^{+99}_{-113}$&$46/2$\\
\hline
\end{tabular}
\caption{Model I - Fe K Gaussian absorption profile components for Suzaku XIS 2013 data. $^t$ denotes that the parameter is tied over the entire sequence 2013a (slices A - D).} 
\vspace{-5mm}
\begin{threeparttable}
\begin{tablenotes} 
	\item[a] Rest-frame energy of the Gaussian absorption line,
	\item[b] equivalent width for a constant width of $\sigma=282_{-56}^{+69}$ eV,
	\item[c]  Change in $\Delta\chi^{2}/{\Delta\nu}$ when the Gaussian component modelling the iron K absorption profile is removed.
	
\end{tablenotes}
\end{threeparttable}
\label{tab:fek_gaussian}
\end{table*}

\subsubsection{\xstar Modelling}

\begin{figure*}


\includegraphics[scale=0.60]{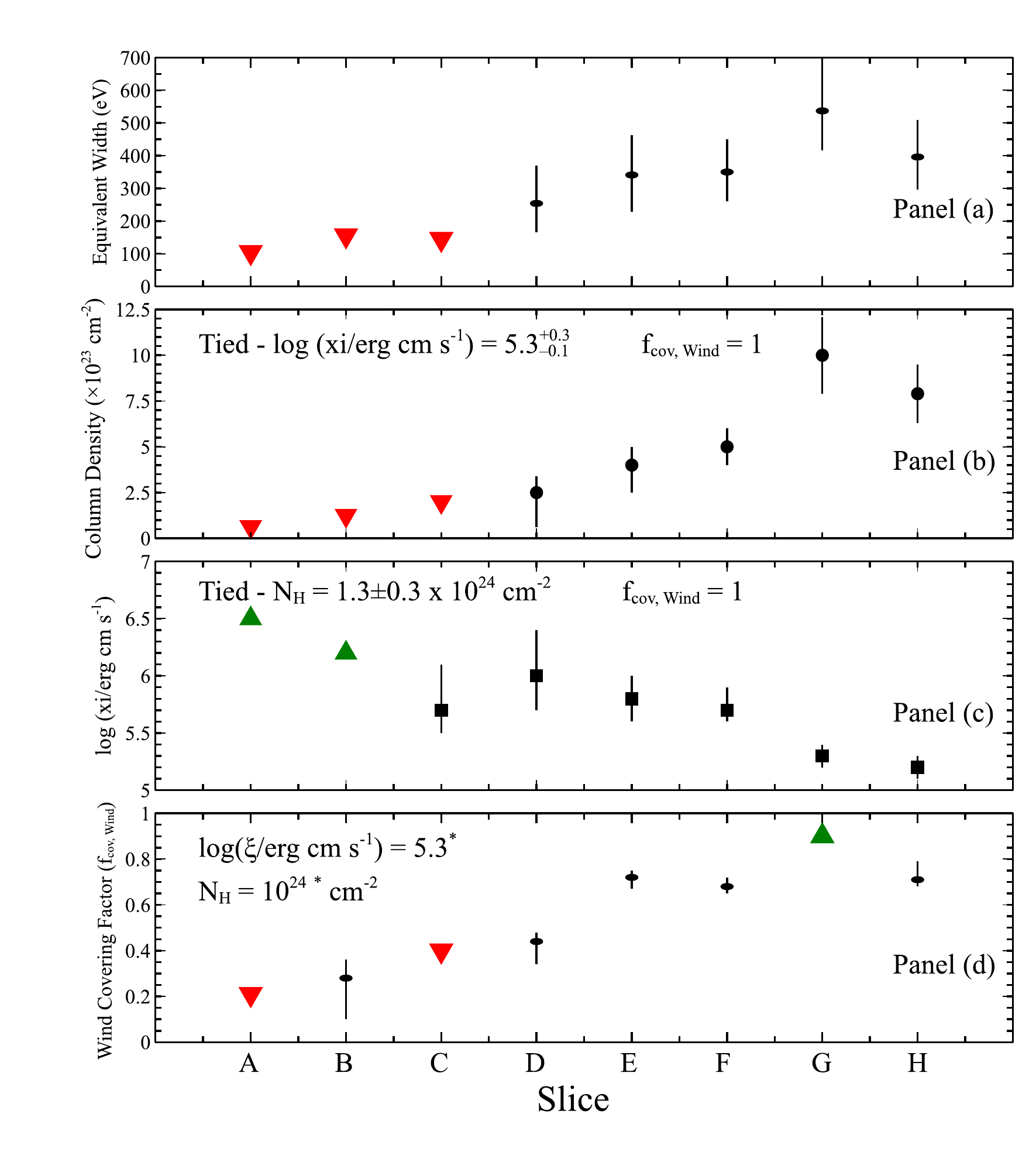}	
\caption{Fe K absorption short-term variability across the observation when modelled with either a Gaussian profile (panel a) or \xstar (panels b - d). (a) EW evolution with time. (b) Column density variability assuming constant ionisation and $f_{\rm cov,Wind}=1$. (c) Ionisation variability assuming constant column density and $f_{\rm cov,Wind}=1$. (d) Variability of $f_{\rm cov,Wind}$ by assuming fixed ionisation and column density.}
\label{fig:pds456_3panels_nh_xi_wind}



\end{figure*}

We then replaced the Gaussian absorption line profile in Model I above with a multiplicative \textsc{xstar} absorption grid, generated as described earlier in Section 4 (hereafter this is referred to as Model II). Three scenarios were investigated to explain the iron K short-term variability, where: (i) the column density of the fully covering wind is allowed to vary, while the ionisation parameter is tied between the slices; (ii) the ionisation parameter of the fully covering wind is allowed to vary, while the column density is tied; (iii) the wind covering factor ($f_{\rm cov,Wind}$) is allowed to vary, while both column density and ionization are fixed across the whole observation. All the \xstar parameters in modelling the Fe K absorption profile are summarised in Table~\ref{tab:xstar_parameters}.
\\ 
Fig.~\ref{fig:pds456_3panels_nh_xi_wind}(b) shows the column density variability, corresponding to case (i) above, by keeping the ionisation constant between the slices at $\log (\xi/{\rm erg~cm~s^{-1}})=5.3_{-0.1}^{+0.3}$. Remarkably, when we compare two adjacent slices the variability is consistent within the errors, but overall the column density of the highly ionised wind increases by a factor of $\sim10$ as the observation progresses -- i.e. from $\log(N_{\rm H}/{\rm cm^{-2})}<22.8$ in slice A to $\log(N_{\rm H}/{\rm cm^{-2})}=24.0\pm0.1$ in slice G. In case (ii), Fig.~\ref{fig:pds456_3panels_nh_xi_wind}(c), with the column density tied between the slices at $\log(N_{\rm H}/{\rm cm^{-2})}=24.1\pm0.1$, we find that instead the ionisation parameter of the wind decreases during the observation reaching its lowest value in slice H, $\log (\xi/{\rm erg~cm~s^{-1}})=5.2\pm0.1$, compared to a lower limit of $\log (\xi/{\rm erg~cm~s^{-1}})>6.5$ in slice A. Thus in this scenario the ionisation varies by a factor of $\sim10$ or more between the two ends of the observation. In physical terms, however, for the ionisation to follow such a trend, the ionising flux has to vary by a factor of $\sim10$, unless the density and/or distance of the gas from the source also varies.\footnote{This latter case is not easily distinguishable from a simple column density change, which is already considered in case (i).} While both cases (i) and (ii) provided an excellent fit to the data, yielding $\chi_{\nu}^{2}=1552/1685$ and $\chi_{\nu}^{2}=1543/1682$ respectively, case (ii) appears to be physically ruled out, as such drastic ionisation changes do not occur in response to similar changes of the hard X-ray continuum. 
\\
We explored case (iii) in Fig.~\ref{fig:pds456_3panels_nh_xi_wind}(d) where, by assuming a fixed column density and ionisation at $\log(N_{\rm H}/{\rm cm^{-2})}=24.0$ and $\log (\xi/{\rm erg~cm~s^{-1}})=5.3$, the variability of the absorption profile is instead accounted for by changes in the covering fraction of the iron K absorber. In this model, in order to compensate for the loss of flexibility as both $N_{\rm H}$ and $\log (\xi)$ are now constant, we also allowed the wind velocity to vary over the single slices. This better accounts for the moderate redwards shift of the line energy in the second part of the observation (see also Table~\ref{tab:fek_gaussian}). We find that the X-ray source is virtually unobscured in slice A (as the ionised absorber is covering $<21\%$), and as the observation progresses the gas moves across the line of sight almost fully covering the source in slice G ($>89\%$ covering); by the end of the observation, the X-ray source may begin to uncover again in slice H ($71_{-3}^{+8}\%$). Thus either $N_{\rm H}$ or $f_{\rm cov,Wind}$ changes may account for the variability of the iron K absorption feature. The physical implications will be examined in detail in the discussion.

\begin{table*}
\begin{tabular}{ccccccccc}

\hline

&A&B&C&D&E&F&G&H\\

\hline

\multicolumn{9}{c}{\textbf{Case (i)} ($\log (\xi/{\rm~erg~cm~s^{-1}})=5.3_{-0.1}^{+0.3}$), $f_{\rm cov,Wind}=1$,$v_{\rm w}=0.245^{+0.003}_{-0.005}c$}\\    

\\

$\log (N_{\rm H}/{\rm cm^{-2}})$  &$<22.8$     &$<23.1$     &$<23.3$     &$23.4_{-0.6}^{+0.2}$    &$23.6_{-0.2}^{+0.1}$   &$23.7\pm0.1$   &$24.0\pm0.1$   &$23.9\pm0.1$\\

\hline

\multicolumn{9}{c}{\textbf{Case (ii)} ($\log (N_{\rm H}/{\rm cm^{-2}})=24.1\pm0.1$), $f_{\rm cov,Wind}=1$,$v_{\rm w}=0.257_{-0.010}^{+0.011}c$}\\

\\

$\log (\xi/{\rm~erg~cm~s^{-1}})$		&$>6.5$  &$>6.2$    &$5.7_{-0.2}^{+0.4}$  &$6.0_{-0.3}^{+0.4}$   &$5.8\pm0.2$     &$5.7_{-0.1}^{+0.2}$   &$5.3\pm0.1$     &$5.2\pm0.1$\\        

\hline 

\multicolumn{9}{c}{\textbf{Case (iii)} ($\log(N_{\rm H}/{\rm cm^{-2}})=24.0^{*}$, $\log (\xi/{\rm~erg~cm~s^{-1}})=5.3^{*}$)}\\

\\

Wind $f_{\rm cov}$     &$<0.21$	&$0.28_{-0.18}^{+0.08}$	 &$<0.40$      &$0.44_{-0.10}^{+0.04}$		&$0.72_{-0.07}^{+0.03}$	&$0.68_{-0.03}^{+0.04}$	   &$>0.89$	   &$0.71_{-0.03}^{+0.08}$\\

$v_{\rm w}(c)$ & $0.24^{t}$ & $0.24^{t}$ & $0.24^{t}$ & $0.24\pm0.02$ & $0.27\pm0.02$ & $0.27\pm0.01$ & $0.24\pm0.01$ & $0.25\pm0.02$\\

\hline
\hline
\end{tabular}
\caption{Model II - \xstar photoionisation model components of the Fe K absorption profile for Suzaku XIS 2013 data. $^{*}$ denotes that the parameter has being fixed, $^{t}$ denotes that the parameter is tied to the first segment value. The red triangles represent upper-bound measurements, while green triangles indicate lower-bound measurements. See the text for the details of the three different cases.} 
\vspace{-5mm}
	

\label{tab:xstar_parameters}
\end{table*}

\subsection{What Causes The Continuum Short-Term Spectral Variability ?}

In addition to the iron K absorption changes, the 2013 observations also show a significant broadband continuum spectral variability. Fig.~\ref{fig:pds456_2013_3lc_1sr} (bottom panel) shows the normalised softness ratio (NSR), computed as the difference between the $0.5-1$ keV over $2-5$ keV softness ratio (as a function of time) and its mean value, divided by the mean itself. This was defined mathematically as ${\rm NSR}=\frac{{\rm SR(t)}-\left \langle {\rm SR} \right \rangle}{\left \langle {\rm SR} \right \rangle}$, and shows the fractional change in the softness ratio (see Fig.~\ref{fig:pds456_2013_3lc_1sr} lower panel). This can be seen in particular prior to and during the course of the flare (slices C and D). This variability can also be appreciated from the spectral shape of the individual slices plotted in Fig~\ref{fig:pds456_eeuf_gamma2_8slice} (top panels).
\\
In particular, we want to test whether the broadband spectral variability is mainly produced by either (i) rapidly varying partial covering absorption or (ii) variations in the intrinsic shape of the continuum, such as the power law and soft excess, while the partial covering parameters ($N_{\rm H},f_{cov}$) are assumed to remain constant between the slices. We adopt MODEL II (including the \xstar modelling of the iron K absorption) as our baseline continuum model to test both scenarios.

\subsubsection{Partial Covering Variability}

In this scenario, the X-ray photons are reprocessed due to the presence of compact clouds of gas, that partially absorb the AGN emission allowing a fraction ($1-f_{\rm{cov}}$) to emerge unattenuated. The size scale of these clouds is typically similar to the X-ray emitting region, of the order of a few tens of $R_{g}$ \citep[e.g.,][]{Risaliti07}. For simplicity, the column densities of the two required partial covering zones are not allowed to vary between the slices, so that the spectral changes over the course of the observation are only due to variations in the covering fractions. Note that we would obtain statistically equivalent results by letting the column densities, rather than the covering fractions, vary between the slices. The relative flux normalisations of the blackbody and the power-law continuum are \textit{only} allowed to vary together by the same factor, i.e. we assumed that there is no intrinsic continuum spectral variability (hereafter Model IIa).
\\
For the partial coverer, the column density of the two regions are found again to be $\log(N_{\rm H,\rm high}/{\rm cm^{-2}})=23.2\pm0.1$ and $\log(N_{\rm H,\rm low}/{\rm cm^{-2}})=22.3\pm0.1$. For the high column zone, the covering fraction reaches its minimum during the flare (slice D) at $f_{cov,\rm high}=0.47_{-0.11}^{+0.09}$, and increases to its maximum value $f_{cov,\rm high}=0.72_{-0.05}^{+0.04}$ in slice E. Thus slice D is the highest flux/least absorbed spectrum overall, while slice E is the lowest flux/most absorbed spectrum. Conversely, the variations in the covering fractions for the low column partial coverer are smaller, simply ranging from $f_{cov,\rm low}=0.72_{-0.07}^{+0.05}$ in slice C to $f_{cov,\rm low}=0.85\pm0.03$ in slice D (see Table \ref{tab:partial_covering}).
\\
This suggests that the high column covering fractions account for most of the spectral variability, and tend to vary of $\pm15\%$ around the mean ($\sim60\%$) but never reaching zero. The range of variability of $f_{cov,\rm high}$ can give us an idea of the possible number of clouds (N) crossing our line of sight at any time, as in first approximation it is related to $1/\sqrt{\rm N}$. This suggests that a large number of clouds (N $\gtrsim25$) are needed along the line of sight to produce small $(\sim20\%)$ variations in covering. However, N$>25$ would imply that the size of each cloud is $\sim1~R_{\rm g}$, thus favouring an alternative scenario where the clouds are just a few but have an irregular or filamentary shape. The temperature of the blackbody component is  $kT=69_{-21}^{+24}$ eV, slightly lower but still consistent with the values previously obtained. Overall, the variable partial covering model provided an excellent fit to the data, with $\chi_{\nu}^{2}=1581/1689$.

\begin{table*}
\begin{center}

\begin{tabular}{cccccccccccc}
\hline
&&\multicolumn{2}{c}{\texttt{\bf power law}} &\multicolumn{2}{c}{\texttt{{\bf bbody}}}&\textbf{\texttt{pc$_{\rm low}$}}& \textbf{\texttt{pc$_{\rm high}$}}&\multicolumn{2}{c}{\textbf{Model Statistic}}  \\

& &\multicolumn{2}{c}{$\Gamma=2.5\pm0.2$}&\multicolumn{2}{c}{${\rm kT}=69_{-21}^{+24}$ eV}&$\log(N_{\rm{H,\rm low}}/{\rm cm^{-2}})$& $\log (N_{\rm{H,\rm high}}/{\rm cm^{-2}})$&\multicolumn{2}{c}{}  \\

\multicolumn{2}{c}{} &\multicolumn{2}{c}{}&\multicolumn{2}{c}{}&$22.3\pm0.1$& $23.2\pm0.1$&\multicolumn{2}{c}{}  \\

\hline

Slice&Time (ks)&norm$^{a}_{\rm po}$&F$_{\rm 2-10}^{b}$&norm$_{\rm bb}^{c}$&F$_{\rm 0.5-2}^{d}$&$f_{cov,\rm low(\%)}$&$f_{cov,\rm high (\%)}$&$(\chi^{2}/{\nu})^{e}$&N.P.$^e$\\

\hline

A&0-100&  $3.3_{-0.8}^{+1.2}$&$2.3$        &$6.0$&     $0.33$&$75_{-6}^{+4}$&$59_{-7}^{+6}$    &$218/224$   &$0.59$\\

B&100-300& $2.4_{-0.5}^{+0.9}$&$1.6$	      &$4.3$ &    $0.18$&$78_{-5}^{+4}$&$66_{-6}^{+5}$    &$241/255$   &$0.73$\\

C&300-400&$2.5_{-0.6}^{+1.0}$&$1.8$        &$4.6$ &    $0.28$&$72_{-7}^{+5}$&$60_{-8}^{+7}$    &$139/143$   &$0.58$\\

D&400-450 &$5.1_{-1.1}^{+1.9}$&$3.9$       & $9.1$&    $0.41$&$85\pm3$&  $47_{-11}^{+9}$       &$148/160$   &$0.75$\\

E&850-1050&$2.1_{-0.5}^{+0.8}$&$1.3$       & $3.8$&    $0.10$&$82_{-5}^{+4}$&$72_{-5}^{+4}$    &$168/193$   &$0.90$\\

F&1050-1250 &$3.0_{-0.7}^{+1.2}$&$1.9$     &$5.4$ &    $0.14$&$84_{-4}^{+3}$&$69\pm5$          &$283/278$   &$0.41$\\

G&1250-1350   &$2.2_{-0.5}^{+0.9}$&$1.6$   &$4.0$ &    $0.14$&$84_{-4}^{+3}$&$61_{-8}^{+6}$    &$128/123$   &$0.36$\\

H&1350-1510 &$2.4_{-0.5}^{+0.9}$&$1.8$     &$4.4$ &    $0.21$&$80_{-4}^{+3}$&$55_{-8}^{+7}$    &$256/243$   &$0.27$\\
\hline

&&&&&$\chi^{2}/{\nu}=1581/1689$&&&\\

\hline

\end{tabular}

\caption{Model IIa parameters for Suzaku XIS 2013 slices. The spectral changes are accounted for by variability of the partial covering absorber. 
The blackbody and power-law normalisations are varying together through the same scale factor throughout the eight slices.} 
\vspace{-5mm}
\begin{threeparttable}
\begin{tablenotes} 
	\item[a] Power-law normalisation, in units of $10^{-3}$ photons keV$^{-1}$ cm$^{-2}$ s$^{-1}$ at $1$ keV, 
	\item[b] observed (non absorption corrected) power-law flux between $2-10$ keV, in units of $10^{-12}$ erg cm$^{-2}$ s$^{-1}$,
	\item[c] blackbody normalisation in units of $10^{-4}$ $(L_{39}/D^{2}_{10})$, where $L_{39}$ is source luminosity in units of $10^{39}$ erg s$^{-1}$ and D$_{10}$ is the distance to the source in units of 10 kpc
	\item[d] observed blackbody (soft excess) flux between $0.5-2$ keV, in units of $10^{-12}$ erg cm$^{-2}$ s$^{-1}$,
	\item[e] $\chi^{2}$, degrees of freedom and null hypothesis probability (N.P.) calculated in each individual slice.

\label{tab:partial_covering}

\end{tablenotes}
\end{threeparttable}

\end{center}
\end{table*}

\begin{table*}
\begin{tabular}{ccccp{1.5cm}ccccc}
\hline
&    &    \multicolumn{2}{c}{\texttt{{\bf power law}} ($\Gamma=2.5\pm0.1$)}   &          \multicolumn{2}{c}{\texttt{{\bf bbody}} (kT$ =69$ eV)}  &       \multicolumn{3}{c}{\textbf{Model Statistic}}&\\
\hline

Slice&Time (ks)&norm$_{\rm po}$&F$_{\rm 2-10}$&norm$_{\rm bb}$&F$_{\rm 0.5-2}$&$(\chi^{2}/\nu)$&N.P.&$(\Delta \chi^{2}/\Delta \nu)$\\
\hline
A&0-100      &$3.6_{-0.9}^{+1.2}$      &$2.5$         &$9.5_{-3.9}^{+4.5}$    &$0.38$   &$271/226$   &$2.0\times10^{-2}$  &$53/2$\\

B&100-300    &$2.3_{-0.6}^{+0.7}$      &$1.6$         &$5.1_{-2.9}^{+2.3}$     &$0.21$   &$244/257$   &$7.2\times10^{-1}$  &$3/2$\\

C&300-400    &$2.8_{-0.7}^{+0.9}$      &$1.9$         &$17.0_{-5.5}^{+7.1}$   &$0.69$   &$177/145$   &$3.5\times10^{-2}$  &$38/2$\\

D&400-450    &$6.0_{-1.5}^{+1.9}$      &$4.1$         &$<14.0$                &$<0.25$   &$170/162$   &$3.0\times10^{-1}$  &$22/2$\\

E&850-1050   &$1.4_{-0.4}^{+0.5}$      &$1.1$         &$<1.4$                 &$<0.06$     &$224/195$   &$7.9\times10^{-2}$  &$56/2$\\

F&1050-1250  &$2.5_{-0.6}^{+0.8}$      &$1.7$         &$<3.6$                 &$<0.15$  &$372/280$   &$1.9\times10^{-4}$  &$89/2$\\

G&1250-1350  &$2.2_{-0.6}^{+0.7}$      &$1.5$         &$<4.5$                 &$<0.18$  &$133/125$   &$3.0\times10^{-1}$  &$5/2$\\

H&1350-1510  &$2.6_{-0.6}^{+0.8}$      &$1.8$         &$13.0_{-3.8}^{+4.9}$   &$0.53$   &$257/245$   &$2.9\times10^{-1}$  &$1/2$\\
\hline
&&&&$\chi^{2}/{\nu}=$1848/1697&&&&\\
\hline
\end{tabular}
\caption{Model IIb parameters for Suzaku XIS 2013 slices for the intrinsic changes scenario. Here the power-law and blackbody components are allowed to vary independently, while $N_{\rm H}$ and $f_{\rm cov}$ are constant for both absorbers. The temperature of the blackbody component was fixed to the best value found in MODEL IIa, i.e.  $kT=69$ eV, in order to avoid physically unrealistically high values to model the curvature present at higher energies. All the quantities and units are the same as in Table~\ref{tab:partial_covering}. The last column shows the difference in $\chi^{2}$ and degrees of freedom compared to Model IIa.} 
\vspace*{-5mm}
\label{tab:table_intr}
%
\end{table*}

\subsubsection{Intrinsic Spectral Variability}

\begin{figure*}
\centering	
\includegraphics[scale=0.35]{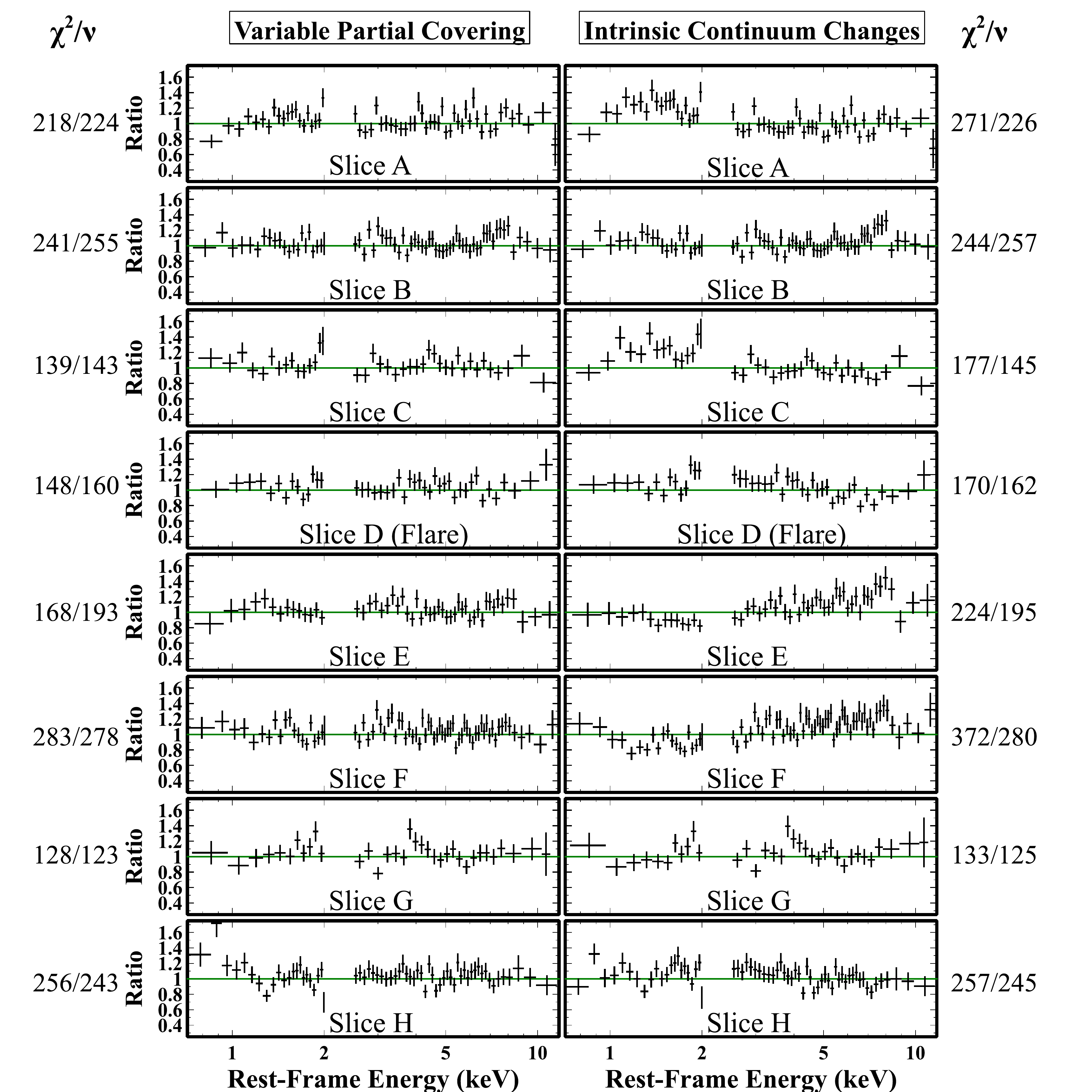}	
\caption{Plots comparing the residuals of each individual slice fitted with a variable partial covering model (left) or intrinsic spectral variability (right). Statistically speaking the $\chi^{2}/\nu$ values of the variable partial covering model are equal or lower in all the eight slices compared to the intrinsic variability fit (see Tables~\ref{tab:partial_covering} and \ref{tab:table_intr}). This is due to the extra curvature present in the residuals when the covering fraction is not allowed to vary.}
\label{fig:2013_pc_vs_intr_8slice_new}
\end{figure*}

Alternatively, we tested if the main driver of the short-term spectral variability is actually a change in the intrinsic continuum via independently variable normalisations of the blackbody and power-law components. The partial covering fractions $f_{cov,\rm low}=81_{-4}^{+3}\%$ and $f_{cov,\rm high}=63_{-6}^{+5}\%$ and column densities $\log(N_{\rm H,\rm low}/{\rm cm^{-2}})=22.3\pm0.1$ and $\log(N_{\rm H,\rm high}/{\rm cm^{-2}})=23.2\pm0.1$ are assumed to be constant between the slices (See Table~\ref{tab:table_intr}). Notably, the values of both $f_{cov,\rm low}$ and $f_{cov,\rm high}$ are almost coincident with the average values found in the previous case. Statistically, this model (hereafter model IIb) gives an overall worse fit ($\chi^{2}/{\nu}=1848/1697$) compared to the variable partial covering scenario, by $\Delta \chi^2/\Delta \nu \sim270/8$. On this basis, we investigated these differences through the inspection of the residuals in each of the eight slices (see Fig.~\ref{fig:2013_pc_vs_intr_8slice_new}).

\noindent Generally speaking, when the partial covering absorption is constant the $\chi^{2}$ is worse in all of the slices, in particular in some of the pre- and post-flare (e.g., C, E, and F). Without allowing the absorber covering fractions to vary, there is in fact an additional curvature present in the residuals as shown in the right panels of Fig.~\ref{fig:2013_pc_vs_intr_8slice_new}. Model IIb seems to account for some of the slices (i.e., slice B, G, and H), although a caveat is that the blackbody normalisations are strongly varying between the slices (e.g., from slice C to E and from slice G to H) and not always in sync with the power law changes, possibly indicating that such extreme intrinsic changes are physically unrealistic (see Table~\ref{tab:table_intr} for values).

\begin{figure}
\centering	
\includegraphics[scale=0.55]{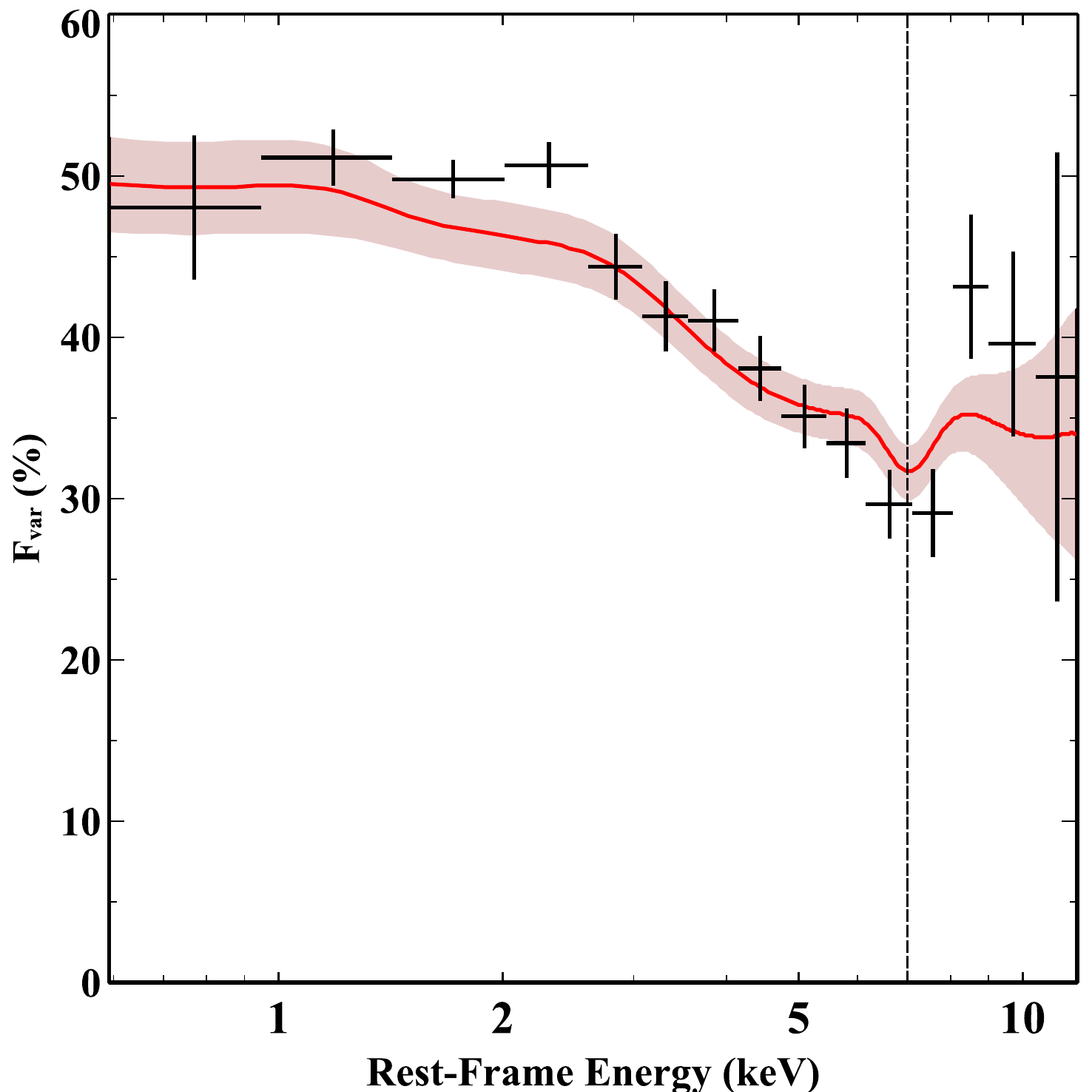}	
\caption{X-ray fractional variability from the 2013 \suzaku observations (in black). The iron K emission band centred at 7 keV (dashed black line) appears to be less variable than the rest of the spectrum. This may be associated with the reprocessed emission arising from the more distant material within the wind. The solid red line corresponds to the simulated F$_{\rm var}$ shape based on MODEL IIa (see text for details). The curve has been smoothed with a spline function, and the shaded area indicates the $1\sigma$ dispersion for $2000$ mock light curves.}
\label{fig:fvar_pds456_2013_2}
\end{figure}

\subsection{Fractional Variability}

In order to further quantify the spectral variability, we also calculated the fractional variability ($F_{\rm var}$) in different energy bands, adopting a time binning of $5760$ s (i.e. one \suzaku orbit) and using the method described in \citet{Vaughan03}. Qualitatively, the overall shape of the broadband $F_{\rm var}$ spectrum (plotted in Fig.~\ref{fig:fvar_pds456_2013_2}) appears broadly consistent with the results of either Model IIa (variable partial covering, Table~\ref{tab:partial_covering}) or Model IIb (intrinsic continuum changes, Table~\ref{tab:table_intr}). Indeed if, the continuum were variable without any spectral variability, which would be the case if the power-law and soft excess components varied in proportion together, then the $F_{\rm var}$ spectrum would, in principle, appear to be constant across all energies. Here instead the variability in the softer band is enhanced with respect to the hard X-ray band. This behaviour could be naturally attributed to: (i) absorption variability seen as partial covering fraction changes across the observation and/or (ii) a two component continuum, where the soft X-ray band component is more variable compared to the hard one. In either case, the soft X-rays would be more affected with respect to the harder X-rays, resulting in an enhanced $F_{\rm var}$ towards lower energies. 
\\
To quantitatively test the $F_{\rm var}$ spectrum against the actual spectral models, we simulated $2000$ light curves based on the spectral parameters obtained for the slices from the previous section. We used as an input for the simulations the best-fitting spectral model of Table~\ref{tab:partial_covering} (MODEL IIa), where the spectral variability is produced by the partial covering absorber. Informed by the best fit parameters from the individual spectral slices, the partial covering fractions ($f_{\rm cov,low}$ and $f_{\rm cov,high}$) were allowed to vary within the ranges from $0.70<f_{\rm cov,low}<0.85$ and $0.4<f_{\rm cov,high}<0.8$. The soft (blackbody) and hard X-ray (power-law) components of the continuum were allowed to vary together (as per MODEL IIa) by a factor of $0.5-2$ in respect to the average normalisation, which represents the expected range of intrinsic continuum variability (see Fig.~\ref{fig:pds456_2013_3lc_1sr}). Each simulated light curve was then randomly generated by varying the spectral parameters within these ranges and the subsequent $F_{\rm var}$ calculated versus energy from averaging over all 2000 simulated curves. 
\\
The result is also shown in Fig.~\ref{fig:fvar_pds456_2013_2}, where the red solid line corresponds to the simulated $F_{\rm var}$ model (and associated $1\sigma$ dispersion shown as the shaded area), compared to the actual $F_{\rm var}$ spectrum measured from the observations. The overall observed shape and normalization of the $F_{\rm var}$ spectrum is  generally well reproduced by the simulations, with the variable partial covering model not only accounting for the rising shape of the $F_{\rm var}$ spectrum towards lower energies, but also for the overall curvature. In contrast, MODEL IIb, where the power-law and blackbody components were allowed to vary independently (with the partial covering absorption held constant), under-predicts the $F_{\rm var}$ spectrum in the $1-3$ keV range. This is likely due to the fact that the soft X-ray blackbody component only adds sufficient variability below 1\,keV and thus the simulated $F_{\rm var}$ spectrum is then flatter compared to the observations. Thus the overall shape of the $F_{\rm var}$ spectrum appears consistent with the variable partial covering scenario presented in the previous section.
\\
Furthermore, looking at Fig.~\ref{fig:fvar_pds456_2013_2}, there is a minimum in $F_{\rm var}$ in the $6.5-7.5$ keV energy band, which is centred on the ionised Fe K emission. This may also suggest that the iron line is less variable compared to the continuum, or at least is not varying on the same short timescale of the X-ray continuum (of the order of $\sim100$ ks). In this light, we investigated from the spectral slices whether the iron K emission line responded to the variability of the continuum. Two opposite scenarios were investigated; first, we kept the Fe K emission line flux fixed between the slices, then we let the emission line flux to vary in sync with the continuum, so that the line EW is constant. As indicated from the $F_{\rm var}$ spectrum, we can achieve a very good fit in the constant flux scenario. In the second case, a constant EW produced a significantly worse fit by $\Delta\chi^{2}/{\Delta \nu}=100/8$, thus indicating that there is no apparent short timescale correlation between the Fe K line and the continuum flux. These results imply that on $\lesssim100$ ks timescales the iron K emission is less variable than the continuum; this suggests that the iron K emitting region is larger than the typical continuum size inferred, which is of the order of $\sim6-20~R_{\rm g}$ (see below). The line may then originate from the outer regions of the disc or, alternatively, if it is associated with the wind as suggested by N15, its lack of variability would be consistent with an origin at distances $\gtrsim100~R_{\rm g}$ from the black hole.
\\
We finally note that, after the iron K emission, the observed variability reaches again a larger value $F_{\rm var}\sim45\%$ value between $8-9$\,keV, coincident with the energy of the iron K absorption feature. This suggests that the iron K absorption variability, discussed earlier in section~\ref{subsec:The Iron K Short-Term Absorption Variability}, may be responsible for this increase in $F_{\rm var}$. For simplicity, however the properties of the highly ionised wind were not allowed to vary when generating the simulated light curves.

\subsection{Properties Of The Partial Covering Absorber}

\begin{figure*}
\begin{center}
\includegraphics[scale=0.65]{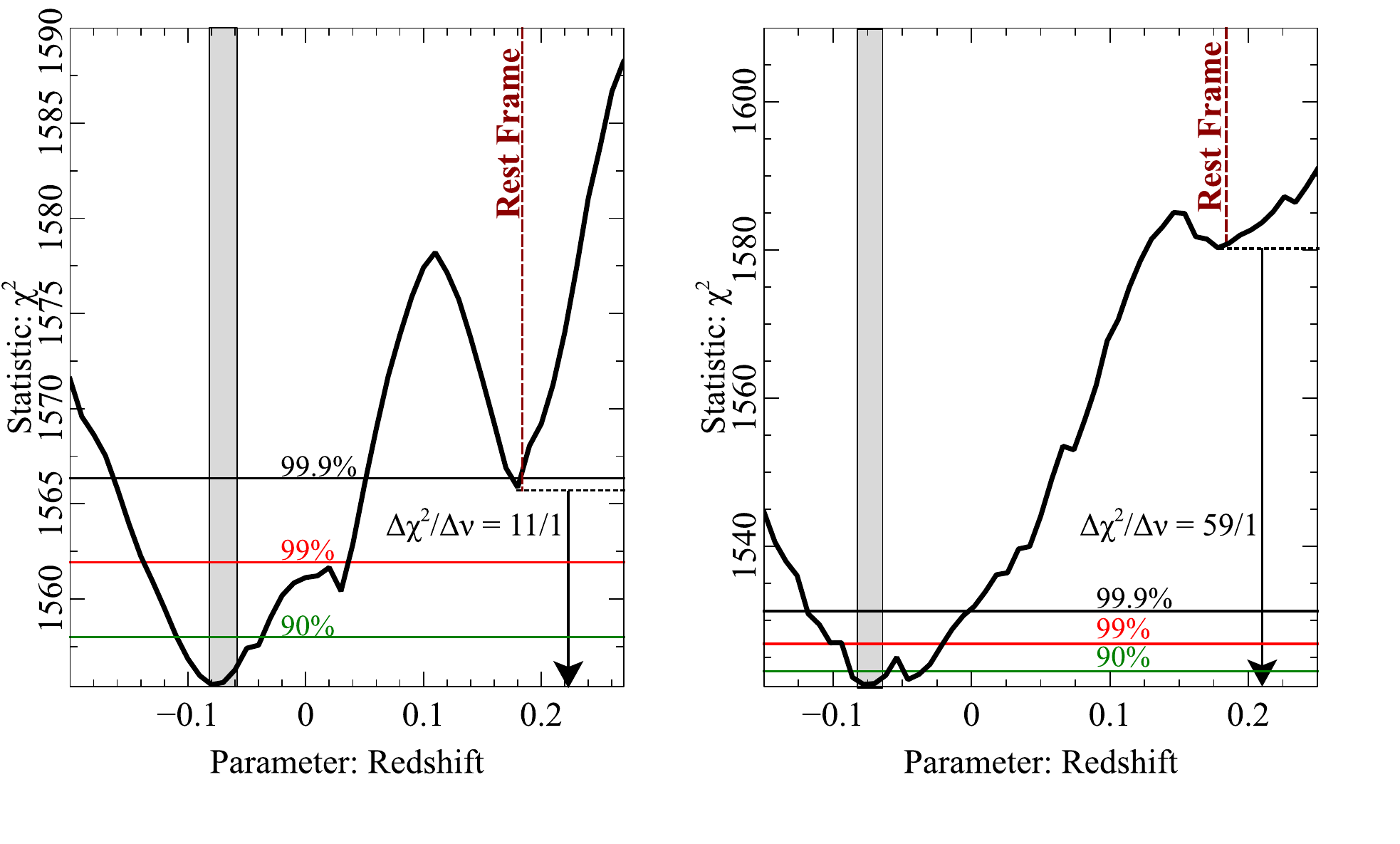}	
\caption{Left: Contour plot of the $\chi^2$ against the partial covering redshift parameter (used as proxy for the velocity) in all the slices. A local minimum at about the quasar rest frame (i.e. $z=0.184$) is clearly visible, but the global minimum indicates that the partial covering layers are possibly outflowing with velocity comparable to that of the fully covering ionised wind, i.e. $v_{\rm pc}\sim~v_{\rm w}\sim0.25c$. This suggests that the two partially covering zones may correspond to the less ionised component of the same ultra fast wind. Right: as for left panel, but with a mildly ionised partial covering absorber. The global minimum corresponds to a statistical improvement, with respect to the rest-frame velocity, of $\Delta\chi^{2}/\Delta{\nu}=59/1$. The vertical shaded area correspond to the range of $v_{\rm w}$ across the slices given in Table~\ref{tab:xstar_parameters}.}
\label{fig:pds456_2013_pc_z_cont}
\end{center}	
\end{figure*}

From the previous sections we conclude that we cannot explain the overall short-term spectral variability without invoking a variable partial coverer. Therefore there is a natural question of what the partial covering is associated with.
\\
One possibility is that it is the less ionised and more dense (or clumpy) part of the outflow. To test this, we investigated the outflow velocity of the partial covering absorber(s) ($v_{\rm pc}$) by allowing it to vary independently from the wind's outflow velocity ($v_{\rm w}$). The resulting fit (applied to the eight slices) suggested that this component is indeed outflowing with a velocity comparable to the velocity of the highly ionised gas. Fig.~\ref{fig:pds456_2013_pc_z_cont} (left) shows the contour plot of the $\chi^2$ against the partial covering redshift parameter tied between all the slices (but allowing the other spectral parameters to vary). A local minimum at about the quasar's rest frame (i.e. $z = 0.184$) is clearly visible, however by assuming a systemic velocity for the partial coverer the fit statistic is worse by $\Delta\chi^{2}/\Delta\nu=11/1$ compared to the best-fit case. The global minimum in Fig.~\ref{fig:pds456_2013_pc_z_cont} indicates that the partial covering gas prefers an outflow velocity of $v_{\rm pc}\sim0.25c$ over a null velocity at the $\sigma=3.29$ ($99.9\%$) confidence level. This indicates that the two partially covering zones might be the least ionised component of the \textit{same} fast $(v_{\rm pc}\sim v_{\rm w}\sim0.25c)$ wind.
\\ 
As a consistency check, we also tested the above result by replacing the two neutral partial covering components with two mildly ionised absorbers, modelled through the same absorption grid responsible for the absorption line at Fe K.
\\ 
We therefore constructed MODEL III expressed as:

\begin{equation}
\begin{split}
\texttt{Tbabs}\times\texttt{xstar}_{\rm Fe}\times\{[(C_{\rm high}\times\texttt{xstar}_{\rm pc,high})\\+(C_{\rm low}\times\texttt{xstar}_{\rm pc,low})+(1-C_{\rm high}-C_{\rm low})]\times\texttt{po}+\texttt{bbody}+\texttt{zgauss}\},   
\end{split}
\end{equation}
\\
where $C_{\rm high}$ and $C_{\rm low}$ are multiplicative constants of value $<1$, \texttt{xstar}$_{\rm Fe}$ represents the fully covering highly ionised wind, while \texttt{xstar}$_{\rm pc,low}$ and \texttt{xstar}$_{\rm pc,high}$ represent the low and high column, mildly ionised partial covering zones respectively. The Gaussian component \texttt{zgauss} again parameterises the ionised emission profile, while the soft excess is described by the \texttt{bbody} component. Note that the model structure is slightly different compared to either Model I or Model II, so that $C_{\rm high}$ and $C_{\rm low}$ are not immediately comparable to the previous $f_{\rm cov,high}$ and $f_{\rm cov,low}$. Here the constants $C_{\rm high}$ and $C_{\rm low}$ physically represent the mutual covering fractions of the high and low column absorbers, so that the remaining fraction $1-C_{\rm high} - C_{\rm low}$ of the continuum is not absorbed by either layer and passes through the partial coverer unattenuated. Also, for simplicity, the blackbody component is assumed to be unobscured. For this reason, its temperature is somewhat larger ($kT=110_{-11}^{+8}$ eV).\footnote{However, by letting the blackbody component vary together with the power law and experience exactly the same absorption, we still achieve a very good fit ($\chi^{2}_{\nu}\simeq0.94$) and the results do not change.} 
\\
Overall, the column densities $\log (N_{{\rm H,high}}/{\rm cm^{-2}})=23.6_{-0.4}^{+0.3}$ and $\log (N_{{\rm H,low}}/{\rm cm^{-2}})=22.5\pm0.3$ are marginally larger (but fully consistent) than in the neutral absorber case, as the gas here is more transparent to the illuminating radiation. We get $\log (\xi/{\rm~erg~cm~s^{-1}})=2.5\pm0.2$ for the high column partial coverer and  $\log (\xi/{\rm~erg~cm~s^{-1}})=0.62_{-0.08}^{+0.18}$ for the low column one. Notably, the high column zone is confirmed to be the main source of the observed continuum variability, and its behaviour is very similar to the one previously described, ranging from $\sim50\%$ (slice D) to $\sim80\%$ (slice E). The fraction of the intrinsic continuum that is not affected by any absorber (apart from the fully covering highly ionised one) is always of the order of a few percent, and never exceeds $10\%$.
\\
The confidence contour plotted in Fig.~\ref{fig:pds456_2013_pc_z_cont} (right) suggests a very similar behaviour in $\chi^{2}$ space for the velocity of the partial coverer. Indeed, the statistical improvement with respect to a systemic velocity is much larger ($\Delta\chi^{2}/{\Delta\nu}=59/1$). Thus we conclude that the partial covering absorbers' outflow velocity, $v_{\rm pc}=0.25_{-0.05}^{+0.01}c$, is consistent with that of the iron K absorber.

\section{Discussion}
\label{sec:Discussion}
\begin{figure*}
\begin{center}
\includegraphics[scale=1.4]{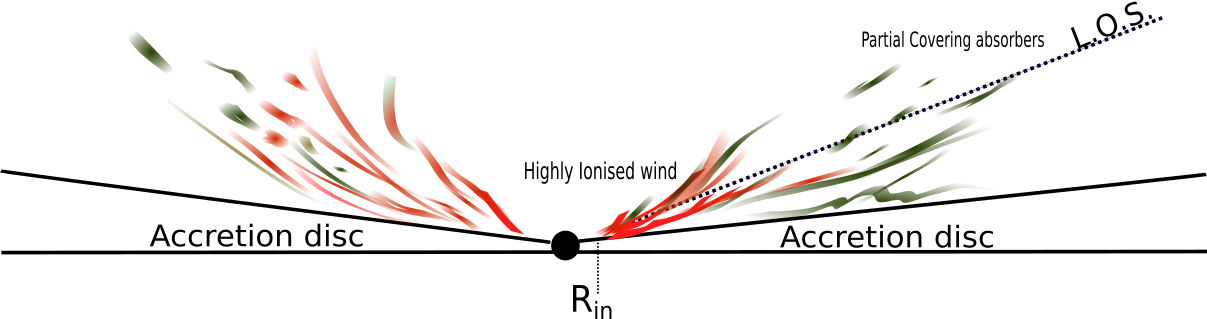}	
\caption{A schematic representation of a possible structure of the outflow and the relative locations of the highly ionised wind (red) and less ionised partial covering absorbers (green) within the outflow. The dashed line represents a possible line of sight. The highly ionised absorber is located in the vicinity of the black hole whilst the less ionised partial covering material is possibly situated further out, both being part of the same outflow.}
\label{fig:wind_pc_cartoon3}
\end{center}	
\end{figure*}

In the previous section we found that the observed short-term spectral variability can be explained by the combination of partial covering absorption and intrinsic continuum variability, although the former effect may be the dominant cause in this observation. Furthermore, by investigating in more detail the properties of the partial covering absorbers, we found that their outflow velocity is comparable to that of the highly ionised wind (i.e. $v_{\rm pc}\sim v_{\rm w}\sim0.25c$) for both the neutral and mildly ionised case. This suggests that the partial covering may be the less ionised but clumpier component of the disc wind, as also supported by the analysis of the soft X-ray absorption features in the high resolution grating \textit{XMM-Newton}/RGS spectra (Reeves et al. 2016, Submitted). A schematic representation of the possible location and structure of the highly ionised wind and partial covering absorbers in context of the outflowing material from the accreting supermassive black hole is shown in Fig.~\ref{fig:wind_pc_cartoon3}. 	

\subsection{Properties of the Clumpy Wind and Constraints on the X-ray Emitting Region}

During the observation, the column density $N_{\rm H}$, or equivalently the EW of the Fe K absorption feature, increased by a factor of $\sim10$, from $\log(N_{\rm H}/\rm cm^{-2})<22.8$ in slice A to $\log(N_{\rm H}/\rm cm^{-2})=24.0_{-0.3}^{+0.1}$ in slice G. Alternatively, these variations may be accounted for by a change in the wind covering factor ($f_{\rm cov,Wind}$), where initially the absorber is not covering the X-ray source but progresses to almost fully covering ($>89\%$) at the time of slice G. These changes could be caused by the transit of a cloud or stream of highly ionised material moving across the line of sight as part of an inhomogeneous outflow (see Fig.~\ref{fig:pds456_3panels_nh_xi_wind}d). The linear size of the transiting clump can be estimated as $\Delta R_{c}\sim v_{\rm w}\Delta t$, assuming that the transverse (Keplerian) velocity is comparable to the outflow velocity (i.e., $v_{\rm K}\sim v_{\rm w}$), and that $\Delta R_{c}$ is similar to the size of the X-ray source. The putative cloud, in fact, cannot be much smaller than the X-ray emitting region in \pds since an almost complete covering is reached in slice G (see Fig.~\ref{fig:pds456_eeuf_gamma2_8slice} bottom panel). At the same time it cannot even be much larger, given that in slice H the covering seems to decrease instead of remaining close to $f_{\rm cov,Wind}\sim1$. We choose $\Delta t=400$ ks, corresponding to the time between the initial (statistically significant, $\sim3\sigma$) onset of the absorption profile in slice E and its completion at maximum depth in slice G ($>5\sigma$). By adopting $v_{\rm w}\simeq0.25c$, we estimate $\Delta R_{c}\approx3\times10^{15}$ cm $\sim20~R_g$ for $M_{\rm BH}\sim10^{9}\Msun$. This can also be seen as a constraint on the size of the X-ray emitting region, which therefore cannot be larger than $\sim20~R_g$ in order for the absorber to reach full covering during slice G. 
\\
For an ionising luminosity of $L_{ion}\sim5\times10^{46}$ erg s$^{-1}$, and taking the average column density between slices E and H of $\log(N_{\rm H}/{\rm cm^{-2}})\sim24$, as well as the average ionisation parameter of $\log(\xi/{\rm erg~cm~s^{-1}})\sim5.3$, the distance of the absorber from the ionising continuum can be estimated. Considering an average hydrogen number density of $n_{\rm H}\sim N_{\rm H}/\Delta R\sim3.3\times10^{8}$ cm$^{-3}$, then from the definition of the ionisation parameter $R=\left ( L_{ion}/n_{\rm H}\xi \right )^\frac{1}{2}\sim2.8\times10^{16}$ cm, or $R\sim180$ $R_{\rm g}$. This radial estimate is of the same order of that measured by N15. These results suggest that we are viewing through a clumpy wind at typical distance of few $100$ $R_{\rm g}$ from the black hole.

\subsubsection{Estimate Of The Wind Radial Distance From Its Keplerian Velocity}

\noindent Above, in estimating the radial distance of the wind from the black hole, we assumed that $v_{\rm K}\sim v_{\rm w}$. Relaxing this assumption, we now instead express the Keplerian velocity as $v_{\rm K}^{2}=\frac{c^2}{r_{\rm g}}$, where $r_{\rm g}=\frac{R}{R_{\rm g}}$ is the distance in units of gravitational radii.

\noindent The hydrogen number density in terms of $r_{\rm g}$ is expressed as

\begin{eqnarray}
n_{\rm H}\sim\frac{\Delta N_{\rm H}}{\Delta R_{c}}\sim\frac{\Delta N_{\rm H}}{\Delta R_{\rm x}}\sim\frac{\Delta N_{\rm H}r_{\rm g}^{1/2}}{c \Delta t},
\end{eqnarray}

\noindent where $\Delta R_{\rm x}$ is the size of the X-ray source. As we practically observe an eclipsing event of the X-ray source, we again assume that $\Delta R_{\rm x}\sim\Delta R_{\rm c}$ \citep{Risaliti07}. $\Delta N_{\rm H}$ is the change in column density between slices E and H of the passing clump as shown in Fig.~\ref{fig:pds456_3panels_nh_xi_wind}b. $\Delta R_{c}=c \Delta t / r_{\rm g}^{1/2}$ is the transverse radial distance covered by the passing clump between slices E - H. 
\\
By substituting the definition of the ionisation parameter we can therefore obtain an expression for the radial distance as,

\begin{eqnarray}
r_{\rm g}^{5/2}=\frac{L_{\rm ion}}{\xi}\frac{ \Delta t~c^5}{\Delta N_{\rm H}}\left (GM_{\rm BH} \right )^{-2}.
\end{eqnarray}

\noindent From the known variables then $R\sim3\times10^{16}$ cm $\sim200 R_{\rm g}$, which is of the same order of the previous estimate. The fast variability likely implies that the wind is not uniform but inhomogeneous. What we see in this observation may be therefore described as a $\sim20$ day long \textquotedblleft picture" of a complex time-dependent AGN accretion disc wind \citep{Proga00,Proga04}, where the line of sight absorption varies over a timescale of days-weeks. On the other hand, G14 investigated whether the increase in depth of the Fe K absorption profile could be attributed to changes in the ionisation state due to recombination within a smooth (constant column density over time) outflow in photo-ionisation equilibrium. In this case, the estimate of the radial extent of the outflow was found to be $R\sim1000$ $R_{\rm g}$. Note, however, that a smooth wind without denser clumps would require another explanation to account for the variable partial covering .

\subsection{Origin and Energetics of the Flare}
\label{Origin and Energetics of the Flare}

In order to analyse the remarkable flare, we focused on the soft ($0.5-1$ keV) and the hard ($2-5$ keV) light curves corresponding to the first $450$ ks of the observation (sequence 2013a, slices A - D). These are respectively plotted in panels 2 and 3 in Fig.~\ref{fig:pds456_2013_3lc_1sr} together with the corresponding softness ratio ${0.5-1}/{2-5}$ keV (panel 4). Here, we observe a steady increase of the soft X-ray flux by a factor of $\sim4$ between $t\sim250-430$ ks. In contrast, the hard X-ray flux shows a sharp increase, by a factor $\sim3$, between $t\sim390-430$ ks. Therefore the doubling time for the soft flux is $\sim90$ ks, whilst the doubling time for the hard flux is $\sim2-3$ times shorter, implying from the light-crossing time argument that the approximate extent of the soft and hard X-ray emission regions in PDS\,456 is $\sim15-20~R_{\rm g}$ and $\sim6-8~R_{\rm g}$ respectively. This suggests a corona characterised by an extended region of \textquotedblleft warm" electrons combined with a more compact region of \textquotedblleft hot" electrons contributing to the rapid hard flare. This would overall be consistent with the earlier representation with the \texttt{optxagnf} disc plus corona model (see section~\ref{sebsec:Modelling The broadband SED}). As we do not have any information just after the flare (i.e., from $t=450$ ks to $t=850$ ks), at this stage we can only provide a speculative interpretation of the possible geometrical relation between the hot and warm coronae. We may either have (i) a layered coronal structure or (ii) a compact hot corona embedded into the warm one. The first case may produce a lag, and cannot be completely ruled out as we are unable to observe the peaks of the hard and soft light curves due to the scheduling gap (see Fig.~\ref{fig:pds456_2013_3lc_1sr}). In the second case, the simultaneous rise of the soft and hard light curves may be interpreted as an accretion rate fluctuation \citep[e.g.,][]{Arevalo06}
with consequent increase of seed disc photons; when the hot corona perceives the increase of the soft X-ray photon density, the observed slope of the hard X-ray light curve significantly steepens. Whatever the physical explanation, this is interestingly consistent with the most recent results from micro-lensing studies, which hint at soft and hard X-ray coronal components of different size \citep{Mosquera13}.  
\\
\noindent One open question involves the origin of the flare itself. A possible scenario is whether the observed flare (in slice D) can be simply due to a \textquotedblleft hole" in the partial coverer, rather than to an intrinsically variable X-ray continuum. On this basis, we tested a model where \textit{only} the covering fractions $f_{cov, \rm low}$ and $f_{cov, \rm high}$ were allowed to vary between the eight slices, whereas the normalisations of blackbody and power law components were not. The overall fit statistic was substantially worse, $\Delta\chi^{2}/{\Delta\nu}=333/14$, in the case where no intrinsic continuum variations were allowed. Thus it is likely that the flare is produced intrinsically.

\subsubsection{Can The Flare Drive The Outflow ?}

\noindent Another question arising is whether the strong flare emission could radiatively power the outflow in the latter part of the observation, as the outflow at iron K is more clearly detected after the occurrence of the flare in the light curve. From the transfer of photon momentum to the wind it follows that:

\begin{eqnarray}
\dot{p}_{\rm w}=\dot{M}_{\rm w} v_{\rm w}\sim\tau\frac{L_{\rm flare}}{c},
\end{eqnarray}

\noindent where $\dot{p}_{\rm w},\dot{M}_{\rm w}$ and $v_{\rm w}$ are the momentum rate, mass outflow rate and outflow velocity of the wind, while the Thomson depth is $\tau\sim1$ for $N_{\rm H}\sim10^{24}~{\rm cm^{-2}}$, as observed in the highly ionised wind.

\noindent Thus the kinetic power (luminosity) of the wind is,

\begin{eqnarray}
\dot{E}_{\rm w}=\frac{1}{2}\dot{M}_{\rm w} v_{\rm w}^{2}\sim\left ( \frac{v_{\rm w}}{2c} \right )L_{\rm flare},
\end{eqnarray}

\noindent or, integrating over time for the total energy:

\begin{eqnarray}
E_{\rm w} \sim\left ( \frac{v_{\rm w}}{2c} \right )E_{\rm flare}.
\end{eqnarray}

For an outflow velocity of $v_{\rm w}\sim0.25c$, then $\dot{E}_{\rm w}\sim\dot{E}_{\rm flare}/8$, hence implying that we would expect only $\sim10-15\%$ of the radiative power in the flare to be directly transferred to the wind.

\noindent From the best fit to the eight slices, we estimated the $1-1000$ Ryd luminosity of the flare to be $L_{\rm Flare}\sim2\times10^{46}$ erg s$^{-1}$. 


\noindent Now, the mass outflow rate of the wind is given by:

\begin{align}
\dot{M}_{\rm{w}} \sim \Omega m_pN_{\rm{H}}v_{\rm{w}}R_{\rm{in}},
\end{align}

\noindent and based on the discussion in the previous sections we have adopted the following variables: $\Omega\sim2\pi$ sr for the solid angle as deduced by N15, the average column density between slices E and H $N_{\rm{H}}>5\times10^{23}$ cm$^{-2}$, $v_{\rm{w}}\sim0.25c$ and $R_{\rm{in}}>100$ $R{\rm_g}\sim1.5\times10^{16}$ cm. Hence we estimated $\dot{M}_{\rm{w}}\ga9\, M_\odot$ yr$^{-1}\sim0.4\dot{M}_{\rm{edd}}$ (in good agreement with N15). The kinetic luminosity of the outflow is $\ga1.5\times10^{46}$ erg s$^{-1}$, or $\sim0.1\, L_{\rm{Edd}}$ for $M_{\rm BH}=10^{9}$\Msun. 
\\
The duration of the wind in slices E to H is at least $\sim600$ ks, thus the mechanical energy carried by  the wind is at least $\sim10^{52}$ ergs. To be conservative, we can assume that the flare is symmetric and persists at its maximum observed luminosity ($L_{\rm Flare}\sim2\times10^{46}$ erg s$^{-1}$) for the entire duration of the observational gap, i.e. for $450$ ks in total. This would impart a total radiative energy of up to $\sim10^{52}$ erg. Indeed, this value represents a maximum upper limit for the total energy radiated during the flare as: (i) it is assumed that the full band $1-1000$ Ryd ionising luminosity has risen in proportion to the X-ray luminosity, which is unlikely be the case; (ii) the flare luminosity is assumed to remain constant for an extended period, contrary to the usual flaring behaviour. Thus the energy radiated during the flare may be an order of magnitude lower if the flare is short lived, which is likely given its rapid rise during slice D. Even for this maximum case, however, given that only $\sim10-15\%$ of the radiative energy is deposited in the wind, it is not possible for the subsequent outflow to be purely driven by the radiation pressure provided by the flare alone. The discrepancy between the observed wind kinetic power and the estimated flare contribution is at least an order of magnitude, and probably even larger. Magnetically driven outflows may instead provide an alternative mechanism for the initial driving for such a powerful wind \citep[e.g.,][]{Ohsuga09,Ohsuga11,Kazanas12,Fukumura10,Fukumura15}. Alternatively the increased opacity through line driving may also be important in PDS\,456 \citep{Hagino15}.

\section{Conclusions}

In this paper we have presented the results from the \suzaku observing campaign of the nearby luminous quasar \pds carried out in early 2013 ($\sim1$ Ms total duration). We investigated the broadband continuum by constructing a spectral energy distribution including the OM and \nustar spectra from the simultaneous \xmm/\nustar campaign in 2013/14. As a result, a physically motivated accretion disc/X-ray corona model like \texttt{optxagnf} was able to account for the optical/UV to X-ray brightness ratio with sensible values of the accretion rate, $\log({\rm L}/{\rm L_{Edd}})\sim-0.1$ and coronal size, $\sim10-40~R_g$. However, this broadband model alone could not account for the spectral curvature present in the low-flux 2013 \suzaku spectra. We show that partial covering absorption is required, and then investigated its variability over the course of the observation. The short-term spectral changes in \pds were interpreted in terms of variable partial covering absorption and/or changes in the intrinsic continuum.
\\
\noindent In the first scenario, we have found that the spectral variability on $\sim100$ ks time-scales may be due to neutral (or mildly ionised) clouds of gas that partially obscure the X-ray source while crossing the line of sight. Statistically speaking, the variable absorber model produces an excellent fit to the data ($\chi^{2}/\nu=1581/1689$). Alternatively, we tested whether the spectral variability can be explained by variations in the intrinsic continuum only, where the partial covering absorbers (still required) are constant throughout the observation. While this model seems to account for some parts of the observation, it fails to account for the spectral variability in particular before and after a prominent flare, during which the flux rises by a factor of $\sim4$ in just $\sim50$ ks. In general we cannot explain the overall short-term behaviour of \pds without invoking a variable partial coverer combined, to a certain extent, with an intrinsically variable continuum. 
\\
\noindent On this basis we have investigated in more detail the properties of the partial covering absorbers. At least two layers of absorbing gas are required, of column density $\log (N_{\rm{H,low}}/{\rm cm^{-2}})=22.3\pm0.1$ and $\log (N_{\rm{H,high}}/{\rm cm^{-2}})=23.2\pm0.1$, with average line of sight covering factors of $\sim80\%$ (with typical $\sim5\%$ variations) and $60\%$ ($\pm10-15\%$), respectively. We have also found that these absorbers may be the least ionised component of the fast wind detected through the Fe\textsc{xxvi} K absorption feature, with typical outflow velocity of $\sim0.25c$. We have shown that the short term variability of the iron K absorption may be attributed to the line of sight variations in the column density (or covering fraction in an X-ray eclipse scenario). Through this variability, the size of the absorber is constrained to be $\sim20~R_{\rm g}$. Following an almost complete obscuration event in the second half of the observation, also the size of the X-ray emitting region cannot be larger than this value. In addition to this, we estimated the typical radial distance of the high ionisation absorber from the black hole to be of the order of $\sim 200$ $R_{\rm g}$. 
\\
\noindent We finally analysed the behaviour of the soft ($0.5-1$ keV) and hard ($2-5$ keV) band light curves during the course of the flare. From the doubling time in flux observed in both the soft and hard band, the X-ray emitting corona may be characterised by a \textquotedblleft warm", extended soft X-ray region of $\sim15-20~R_{\rm g}$, which also includes a more compact ($\sim6-8~R_{\rm g}$) zone of \textquotedblleft hot" electrons, responsible for the rapid rise in the $2-5$ keV light curve. We ruled out that the flaring behaviour is instead due to a sudden \textquotedblleft hole" 
in the inhomogeneous absorber surrounding the X-ray source.
\\
\noindent We have explored whether the radiation pressure imparted by the flare could deposit enough kinetic power in the outflowing material to drive the wind. The estimated radiative power of the flare was less than the mechanical power measured for the outflow by at least one order of magnitude, leading to the conclusion that another physical launching mechanism (e.g., magneto-hydrodynamic) is likely contributing as well. 
\\
\noindent Finally, we calculated the fractional variability in the 2013 dataset as a function of energy. We found that the iron K emission band is somewhat less variable than the rest of the continuum. This suggests that the iron K emission occurs from a much larger region than the continuum. This is consistent with the feature arising from size scales $\gtrsim100~R_{\rm g}$, i.e., within the wind itself.

\section{Acknowledgement}

We thank the anonymous referee for their careful report, which helped us improving the clarity of the paper. This research has made use of data obtained from the \suzaku satellite, a collaborative mission between the space agencies of Japan (JAXA) and the USA (NASA). GM, JR, EN, MC and JG all acknowledge the financial support of STFC. VB acknowledges the support from the grant ASI-INAF NuSTAR I/037/12/0.

\bibliographystyle{mn2e}
\bibliography{gabi_first_paper}

\end{document}